\documentclass[11pt]{article}

\usepackage{acl}

\usepackage{times}
\usepackage{latexsym}
\usepackage[T1]{fontenc}
\usepackage[utf8]{inputenc}
\usepackage{microtype}
\usepackage{inconsolata}
\usepackage{graphicx}
\usepackage{booktabs}
\usepackage{multirow}
\usepackage{amsmath}
\usepackage{amssymb}
\usepackage{xcolor}
\usepackage{algorithm}
\usepackage{algpseudocode}
\usepackage{listings}
\usepackage{xspace}
\usepackage{placeins}
\usepackage{hyperref}

\algrenewcommand\algorithmicrequire{\textbf{Input:}}
\algrenewcommand\algorithmicensure{\textbf{Output:}}

\newcommand{\repot}{\textsc{RePoT}\xspace}
\newcommand{\pot}{\textsc{PoT}\xspace}
\newcommand{\Cot}{\textsc{CoT}\xspace}
\newcommand{\sct}{\textsc{SC}\xspace}

\newcommand{\puzzlezoo}{\textsc{PuzzleZoo}\xspace}
\newcommand{\derail}{\textsc{Derail}\xspace}
\newcommand{\arepot}{\textsc{Adaptive RePoT}\xspace}

\title{\repot: Recoverable Program-of-Thought\\
       via Checkpoint Repair\thanks{Code: \url{https://github.com/parsa-mz/RePot}}}

\author{Parsa Mazaheri \\
        University of California, Santa Cruz \\
        \texttt{pmazaher@ucsc.edu}}

\begin{document}
\maketitle

\begin{abstract}
One-shot Program-of-Thought (\pot) emits a Python program that prints a
primitive-action plan; a single invalid action silently invalidates the
trajectory. We introduce \textbf{\repot} (Recoverable \pot): a
deterministic \emph{verified replay} that walks the plan through the
environment to its first invalid transition, then one LLM call that
resumes from the verified prefix. \repot costs at most one extra LLM
call on the $\sim$$14\%$ of problems where \pot fails. \repot beats
\pot by $+3$ to $+11$pp across four closed-model configurations on
\puzzlezoo-775 and peaks at $96.9\%$ vs $86.3\%$ on
\texttt{gpt-5.4-mini-medium}; against the matched-budget \pot-retry
baseline, \repot wins decisively on Gemini ($+3.8$ pp, $95\%$ CI
$[+2.2,+5.4]$), is within sampling noise on GPT-medium and Claude,
and loses on GPT-mini --- a capability-scaling pattern we begin to
address with \arepot, a rule-based dispatcher that routes between
suffix repair and a fresh \pot retry based on verified-prefix length
(preliminary; App.~\ref{app:opensource_table}). We replicate on
\textsc{PlanBench Blocksworld} ($+1.1$ to $+11.4$pp) and on four
open-weights models ($+3.3$ to $+20.0$pp on three of four). On
\derail-550, our controlled recovery benchmark, every condition with
access to checkpoint information clears $\geq$$30\%$ on GPT-medium and
$\geq$$70\%$ on Gemini, vs $\leq$$3.1\%$ for error-only feedback ---
showing that \emph{checkpoint information}, not the specific
verified-prefix tail, is the load-bearing recovery signal.
\end{abstract}

\section{Introduction}
\label{sec:intro}

Large language models can sketch impressive plans, then commit one
illegal action and silently fail the entire task. The dominant fix is to
either run a single sample through a tool (Program-of-Thought, \pot
\citealp{chen2022pot}) or sample many independent rollouts and aggregate
(Self-Consistency \citealp{wang2022sc}, Tree of Thoughts
\citealp{yao2023tot}). Neither is recoverable: a one-shot \pot plan with a
mid-rollout error cannot resume from where it succeeded; tree-search methods
pay branching cost during generation regardless of whether the first
trajectory was already mostly correct.

We propose \repot --- \emph{Recoverable Program-of-Thought} --- a small
modification of one-shot \pot that adds checkpoint-based recovery without
moving to tree search. \repot works in three steps (Algorithm
\ref{alg:repot}):

\begin{enumerate}
\item Run \pot once: emit a Python program, execute it, parse the
      printed move list.
\item \emph{Verified replay}: walk the proposed actions through the
      environment one step at a time, accumulating a maximal
      \emph{verified prefix} of valid transitions until the first failure
      (Eq.~\eqref{eq:replay}).
\item If the prefix already reaches the goal, return success. Otherwise
      issue \emph{one} suffix-repair LLM call, conditioning on the verified
      prefix, the verified state at the failure boundary, and the
      verifier's error message.
\end{enumerate}

\paragraph{Novelty and positioning.}
\repot reframes one-shot LLM reasoning as a \emph{recoverable
execution}: rather than all-or-nothing, the verifier owns the trusted
state and the model is only ever asked to repair the unverified suffix.
Three pieces compose: (i)~a deterministic verified-replay primitive
that turns any \pot rollout into a checkpoint-resumable computation
with no LLM calls; (ii)~a suffix-repair call conditioned on the
\emph{verified state} rather than a textual critique of the prior
attempt --- one well-typed task; (iii)~a single repair budget
($R\!=\!1$), so cost stays at \pot baseline on the $\sim$$86\%$ of
easy problems and only doubles on the rest. Where Reflexion-style
\citep{shinn2023reflexion} verbal critique asks the model to introspect
its mistakes textually, \repot makes the verifier the source of truth.
Where ToT and LATS \citep{zhou2023lats} branch \emph{during}
generation, \repot branches only \emph{after} a deterministic check
identifies a real failure point. \derail-550
(\S\ref{sec:mechanism}) shows the \emph{trusted checkpoint} is the
load-bearing signal: $80.7\%$ recovery of injected errors versus
$20.7\%$ from error-only feedback.

\paragraph{Contributions.}
\begin{itemize}
\item \textbf{Algorithm.} \repot --- a recoverable extension of \pot
      that combines deterministic verified replay (Eq.~\ref{eq:replay})
      with a single suffix-repair LLM call.
\item \textbf{Empirical.} On \puzzlezoo-775, a verifier-backed suite,
      \repot improves over \pot by $+3$ to $+11$pp across three frontier
      models in four configurations, replicates on \textsc{PlanBench Blocksworld}
      \citep{valmeekam2023planbench} ($+1.1$ to $+11.4$pp), and beats a
      matched-budget \pot-retry baseline on the two reasoning-enabled
      models.
\item \textbf{Mechanism.} A controlled \derail-550 benchmark
      isolates which signal makes recovery work: trusted
      checkpoint state separates the recovery-capable conditions from
      error-only feedback by $\sim$$60$pp; the explicit verified-prefix
      tail provides a smaller, model-dependent additional benefit.
\end{itemize}

\section{Background}
\label{sec:background}

\paragraph{Program-of-Thought planning.}
\pot \citep{chen2022pot} prompts the model to write a Python program
whose printed output is the plan, then runs the program in a sandbox and
parses the printed token list. The original work targeted arithmetic and
symbolic reasoning, where the program is the answer. In LLM planning,
the printed plan is a sequence of primitive actions $(a_1, \ldots, a_n)$
which is then executed by an environment simulator, and the trajectory
either reaches the goal or it does not. \pot's appeal in this setting is
that it pushes the brittle bookkeeping (state tracking, legal-action
filtering, search) onto the model in code form, which the model is good at
\citep{wei2022cot, wang2022sc}. The cost is also its weakness: a single
illegal action mid-rollout invalidates the entire trajectory, and the
trajectory's verified prefix is discarded along with the invalid suffix.

\paragraph{The Illusion-of-Thinking finding.}
\citet{shojaee2025illusion} (``The Illusion of Thinking'') run a controlled
puzzle benchmark across four classical planning environments
(Tower of Hanoi, Checker Jumping, River Crossing, Blocksworld) at
controllable complexity, and report that frontier reasoning models exhibit
sharp accuracy collapse once complexity exceeds a model-specific
threshold, even when given the algorithm. They further show that
failures concentrate early in the trace: the first invalid move
often appears at a small fraction of the optimal-plan length, and the
remaining tokens are spent on a wrong-but-consistent continuation.
\citet{song2025thinking} re-run a subset and argue some collapses are
artifacts of impossible River Crossing instances and prompt encoding;
\citet{scholten2024metacog} frame the same phenomenon as
``metacognitive myopia''. Whatever the framing, the empirical pattern is
robust: most failed traces have a long valid prefix and a single
mid-rollout misstep. This is precisely the regime in which
checkpoint-based recovery should help. \repot is a targeted fix for
this \emph{recoverable subset} of failures: we do not claim to solve
reasoning collapse in general, only to mitigate \emph{recoverable
execution collapse} where a trusted intermediate state can be preserved
and resumed from.

\paragraph{Prior work on one-shot brittleness.}
A large body of work has proposed remedies for the one-shot failure
mode; we group them into three families.
\textbf{(i)~Sample more}: Self-Consistency \citep{wang2022sc} and
best-of-$k$ generate $k$ trajectories and vote / pick the best by some
scoring function. Cost is $O(k)$ LLM calls per problem regardless of
whether the first trajectory was nearly correct.
\textbf{(ii)~Branch during generation}: Tree of Thoughts
\citep{yao2023tot} and LATS \citep{zhou2023lats} treat reasoning as
search, expanding multiple continuations and using a value model or
verifier to prune. They pay tree-search cost on every problem, including
easy ones.
\textbf{(iii)~Iterate with critique}: Reflexion \citep{shinn2023reflexion},
ReAct \citep{yao2023react}, Self-Refine \citep{madaan2023selfrefine}, and
Self-Repair \citep{olausson2024selfrepair} run a second LLM call
conditioned on a textual critique of the prior attempt. The critique can
be wrong (\emph{the model that produced the failure now also produces the
diagnosis}), and there is no checkpoint mechanism: the second call
re-plans from scratch given the critique.

\paragraph{Process verification and rewards.}
A separate line uses process reward models (PRMs)
\citep{lightman2023prm800k} to score partial reasoning traces and
reject low-scored continuations. PRMs need a learned scorer and target
mathematical reasoning where ground-truth verification is hard. \repot
sits in the complementary regime where the verifier is the environment
itself --- exact, free, and immediately available --- which lets us skip
the learned scorer entirely.

\section{Related Work}
\label{sec:related}

Section~\ref{sec:background} surveyed the main families that
\repot relates to. Here we position \repot against three further
adjacencies. \textbf{Concurrent transactional/checkpoint work}
\citep{atomix2026, sagallm2025, safeflow2025} targets multi-agent
coordination rather than one-shot \pot. \textbf{Decoupled
reasoning--observation} (ReWOO, \citealp{xu2023rewoo}; ThinkSwitcher,
\citealp{thinkswitcher2025}) routes between thinking modes per input;
\repot routes per-trajectory, conditioned on a verified failure
boundary. \textbf{Planning benchmarks} \textsc{PlanBench}
\citep{valmeekam2023planbench, valmeekam2024planning} provides
PDDL-grounded LLM planning evaluation; we use its Blocksworld split as
external replication and discuss the agentic-gap framing
\citep{agenticgap2025} in Discussion.

\section{Method}
\label{sec:method}

\begin{figure*}[t]
\centering
\includegraphics[width=0.95\textwidth]{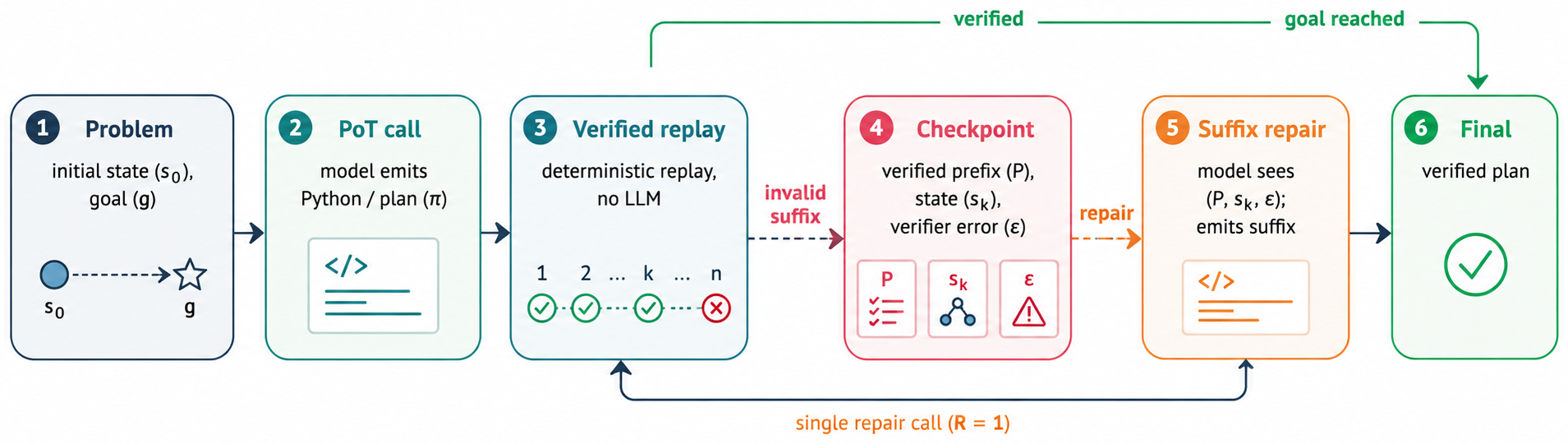}
\caption{The \repot pipeline. \textbf{(1)~Problem} provides initial state
$s_0$ and goal $g$. \textbf{(2)~\pot call} (LLM call \#1): the model
emits a Python program whose stdout encodes the action plan $\pi$.
\textbf{(3)~Verified replay}: walk $\pi$ through the environment one
step at a time --- deterministic, no LLM calls --- producing the
maximal valid prefix and the failure boundary. If every action is
valid, branch to \textbf{(6)~Final} via the upper \emph{verified, goal
reached} arrow. Otherwise the boundary is packaged as
\textbf{(4)~Checkpoint}: verified prefix $P$, verified state $s_k$, and
verifier error $\varepsilon$. The \textbf{(5)~Suffix repair} call (LLM
call \#2) sees the checkpoint and emits a suffix that resumes from
$s_k$. The lower arrow is the single repair budget ($R\!=\!1$): the
suffix is itself sent through verified replay; on success we land at
Final, on failure we accept the partial verified plan.}
\label{fig:schematic}
\end{figure*}

\subsection{Problem setting}
\label{sec:method:setting}

We consider planning tasks of the form $(s_0, g, E)$, where $s_0$ is the
initial state, $g$ is the goal specification, and $E$ is a deterministic
environment with a step function $E\!:\!(s, a) \mapsto (s', \text{valid})$
that returns the next state and a validity flag. A plan is a sequence of
primitive actions $\pi = [a_1, a_2, \ldots, a_n]$. Success is
$\textsc{IsGoal}(s_n, g)$ where $s_n$ is the result of replaying $\pi$
from $s_0$. We assume the environment exposes a verifier (the same
\textsc{Step}) and a goal predicate; we do not assume the model has
direct access to either.

\subsection{Verified replay}
\label{sec:method:replay}

The core primitive is a deterministic \emph{verified-replay} function that
takes a candidate action sequence and walks it through the environment.
Given start state $s_0$ and a plan $\pi = (a_1, \ldots, a_n)$, define the
trajectory $s_i = \mathrm{step}(s_{i-1}, a_i)$ if the transition is valid,
else $\bot$. Let $k$ be the smallest index for which the transition is
invalid (or $n+1$ if every action is valid). Then
\begin{equation}
\textsc{Replay}(E, s_0, \pi) \;=\; \bigl(P,\; s_{k-1},\; \epsilon_k\bigr),
\label{eq:replay}
\end{equation}
where $P = (a_1, \ldots, a_{k-1})$ is the maximal verified prefix,
$s_{k-1}$ is the verified state at the failure boundary, and $\epsilon_k$
is the verifier's error message at step $k$ (or the empty string if
$k = n+1$). The function is total, deterministic, and makes no LLM calls;
its cost is $O(n)$ environment steps.


\subsection{\repot algorithm}
\label{sec:method:repot}

\repot composes \pot with verified replay and a single suffix-repair call.
The full algorithm is shown in Algorithm~\ref{alg:repot}. Two
hyperparameters govern its behaviour: the repair budget $R$ (default
$R\!=\!1$) and the verified-prefix tail length $T$ shown to the model
during repair (default $T\!=\!4$). The model is otherwise sampled at
$\tau\!=\!0$.

\begin{algorithm}[t]
\caption{\repot --- Recoverable Program-of-Thought}
\label{alg:repot}
\begin{algorithmic}[1]
\Require problem $(s_0, g)$, env $E$, model $\mathcal{M}$, repair budget $R$
\State $\pi \gets \mathcal{M}\bigl(\textsc{PoTPrompt}(s_0, g)\bigr)$
       \Comment{1. one-shot \pot}
\State $(P, s, \epsilon) \gets \textsc{Replay}(E, s_0, \pi)$
       \Comment{2. verified replay (Eq.~\ref{eq:replay})}
\If{$\textsc{IsGoal}(s, g)$} \Return $P$ \EndIf
\For{$r = 1, \ldots, R$}
       \Comment{3. suffix repair from checkpoint}
  \State $\pi' \gets \mathcal{M}\bigl(\textsc{RepairPrompt}(s_0, g, s, P, \epsilon)\bigr)$
  \State $(P', s, \epsilon) \gets \textsc{Replay}(E, s, \pi')$
  \State $P \gets P \mathbin{+\!\!+} P'$
  \If{$\textsc{IsGoal}(s, g)$} \Return $P$ \EndIf
\EndFor
\State \Return $P$ \Comment{partial; budget exhausted}
\end{algorithmic}
\end{algorithm}

\paragraph{A simple recovery model.}
For a problem instance, let
$p$ be the probability that \pot succeeds on the first sample,
$q$ the probability that \pot fails but leaves a recoverable valid
prefix, $r$ the conditional probability that suffix repair succeeds
given a recoverable prefix, $b$ the conditional probability that a
fresh \pot resample succeeds given the first sample failed, and
$b'\!\leq\!b$ the fresh-retry success rate restricted to the
unrecoverable subset.
\repot beats \pot-retry iff
\begin{equation}
q\,(r - b) \;>\; (1-p-q)\,(b - b'),
\label{eq:condition}
\end{equation}
i.e.\ when verified-prefix repair beats the fresh-sample marginal on
the recoverable subset. Larger $q$ (longer valid prefixes, scaling with
capability) makes the condition more favourable;
\S\ref{sec:results:opensource} and Fig.~\ref{fig:capability_scaling}
show this empirically. The adaptive
variant below dispatches per-problem to maximize Eq.~\ref{eq:condition}.

\subsection{Adaptive recovery policy}
\label{sec:method:adaptive}

Algorithm~\ref{alg:repot} always commits to the verified prefix. When the
prefix is empty or very short, the verified state $s$ collapses to $s_0$
and the repair call effectively restarts from the initial state but with
a (potentially misleading) error anchor. Empirically, on weaker models
\pot-retry's fresh sample outperforms anchoring on a short prefix
(\S\ref{sec:results:headline}).

As a preliminary extension, we introduce \textbf{\arepot}, a rule-based
dispatcher with the same $R\!=\!1$ budget. After verified replay, we
read the prefix fraction $\phi = (k-1)/n$ and route to a \emph{fresh
\pot retry} when $n\!=\!0$ or $\phi\!<\!0.15$, otherwise to suffix
repair (Alg.~\ref{alg:repot}). Thresholds were fixed a priori (not
tuned on test); a threshold sweep and alternative dispatcher rules are
left to future work. The dispatcher realizes the optimal-branch
prediction implied by Eq.~\ref{eq:condition}: route to fresh sampling
when the recoverable subset is empty, otherwise exploit the verified
prefix. Open-source results are in App.~\ref{app:opensource_table}.

\subsection{Repair prompt and ablations}
\label{sec:method:prompt}

The repair call uses a verified-prefix-conditioned prompt: a stable block
(problem statement, goal) above a verifier-checkpoint marker, and a
dynamic block (last $T$ verified moves, current verified state, legal
actions, verifier error) below. Splitting along this boundary makes the
stable block prefix-cacheable across repair calls; the full template is
in App.~\ref{app:prompts} (Fig.~\ref{fig:repair-prompt-app}).

Three named ablations are used in \S\ref{sec:mechanism}:
\textbf{\repot$_\textnormal{full}$} (Algorithm~\ref{alg:repot}),
\textbf{\repot$_\textnormal{no-prefix}$} (hides the prefix tail; model
sees only the current verified state + error), and
\textbf{\repot$_\textnormal{restart}$} (repair call restarts from $s_0$
instead of the verified $s$, isolating ``checkpoint'' from
``extra call''). Definitions in App.~\ref{app:ablations}.

\section{Experimental Setup}
\label{sec:setup}

\subsection{Benchmarks}
\label{sec:setup:benchmarks}

\paragraph{\puzzlezoo-775.}
A stratified problem set across four classical planning environments:
Tower of Hanoi (8 complexities, 200 problems), Checker Jumping (9
complexities, 225 problems), River Crossing (4 complexities, 100
problems), and Blocksworld (10 complexities, 250 problems). Each
environment exposes \textsc{Step}, \textsc{IsGoal}, \textsc{Normalize},
and \textsc{LegalActions} interfaces.

\paragraph{\textsc{PlanBench} Blocksworld (378 problems).}
We use both subsets of \textsc{PlanBench}'s Blocksworld split
\citep{valmeekam2023planbench}: \texttt{generated\_basic} (189 4-block
instances) and \texttt{generated} (189 instances spanning $3$--$12$
blocks). We adapt the PDDL semantics into our \textsc{Step} interface
(predicate state, four-op action vocabulary $\{$pick-up, put-down, stack,
unstack$\}$, partial-goal subset check); see Appendix~\ref{app:planbench}
for the adapter.

\paragraph{\derail-550 (550 errors $\times$ 11 conditions).}
A controlled recovery-from-injected-error benchmark we build alongside
\puzzlezoo-775. For each problem, we run an oracle plan to a checkpoint
$\sim$$1/3$ of the way through, inject one randomly chosen wrong action,
and ask each recovery method to take over. We compare $11$ conditions
(\S\ref{sec:mechanism}) including \repot's three prefix ablations.

\subsection{Models}
\label{sec:setup:models}

We evaluate \repot on both closed-source and open-weights models. The
closed set comprises three frontier models in four configurations:
\texttt{gpt-5.4-mini-medium} (reasoning \texttt{medium}),
\texttt{gpt-5.4-mini} (no reasoning),
\texttt{gemini-3.5-flash} (\texttt{thinking}=\texttt{MEDIUM}), and
\texttt{claude-sonnet-4.6} (no thinking). The open-weights set
comprises four models served on a single NVIDIA H100 80GB GPU via vLLM
with extended thinking disabled:
\texttt{Qwen3.6-35B-A3B} \citep{qwen2026qwen36},
\texttt{gemma-4-26B-A4B-it} \citep{gemma4_2025},
\texttt{gpt-oss-20b} \citep{openai2025gptoss}, and
\texttt{Nemotron-3-Nano-30B-A3B} \citep{nvidia2025nemotron3nano}.
Sampling is deterministic ($\tau\!=\!0$); full hyperparameters are in
Table~\ref{tab:hyperparams} (App.~\ref{app:hyperparams}).

\subsection{Methods compared}
\label{sec:setup:methods}

\Cot, \pot, Self-Consistency (\sct, $k\!=\!8$), \pot-retry, and our
\repot ($R\!=\!1$, $T\!=\!4$). \Cot and \sct emit prose plans; \pot,
\pot-retry, and \repot all emit Python code. \pot-retry is a
matched-budget control: run \pot once, and on verifier failure run \pot
once more from scratch with no prefix and no checkpoint --- the same
two-LLM-call worst-case budget as \repot, but with no checkpoint
mechanism. The \pot vs \pot-retry vs \repot triple lets us separate
re-rolling from genuine checkpoint-based recovery.

\section{Results}
\label{sec:results}

\subsection{Headline cross-model accuracy}
\label{sec:results:headline}

Table~\ref{tab:headline} reports success rate on \puzzlezoo-775
for the four closed models. \repot beats \pot on every model with
$\Delta \in [+2.7, +10.6]$pp; the largest improvement is on
\texttt{gpt-5.4-mini-medium} ($96.9\%$ vs $86.3\%$). Against the
matched-budget \pot-retry baseline, \repot wins decisively on Gemini
($+3.8$, $95\%$ CI $[+2.2,+5.4]$), is statistically a tie on
GPT-medium and Claude (CIs cross zero), and loses on GPT-mini
($-6.6$, $[-9.4,-3.7]$). The result is consistent with our central
claim that \repot's mechanism contribution scales with the validity
of the first \pot plan, which itself scales with model capability
(\S\ref{sec:discussion}). Per-method
cost is in App.~\ref{app:cost} (Fig.~\ref{fig:pareto}).

\begin{table*}[t]
\centering\small
\setlength{\tabcolsep}{5pt}
\begin{tabular}{lrrrrrr}
\toprule
Model & CoT & SC$_{k=8}$ & PoT & PoT-retry & \textbf{RePoT} & R$-$PR \\
\midrule
\multicolumn{7}{l}{\emph{Closed frontier models}} \\
GPT-5.4-mini (med)\textsuperscript{$\star$}  & 11.6 & 37.5 & 86.3 & 96.6 & \textbf{96.9} & $+0.3$ \\
Gemini 3.5 Flash\textsuperscript{$\star$}    & 83.0 & 96.6 & 81.3 & 84.1 & \textbf{87.7} & $+3.6$ \\
Claude Sonnet 4.6                            & 44.1 & 75.9 & 83.1 & \textbf{87.5} & 86.1 & $-1.4$ \\
GPT-5.4-mini                                 & 17.9 & 24.4 & 58.7 & \textbf{68.0} & 61.4 & $-6.6$ \\
\midrule
\multicolumn{7}{l}{\emph{Open-source models}} \\
Gemma 4 26B-A4B        & 18.3 & 24.2 & 49.2 & 54.2 & \textbf{69.2} & $+15.0$ \\
GPT-OSS 20B            & 31.7 & 58.3 & 48.3 & 59.2 & \textbf{62.5} & $+3.3$ \\
Qwen 3.6 35B-A3B       & 36.7 & 59.2 & 55.0 & \textbf{61.7} & 58.3 & $-3.4$ \\
Nemotron-3 Nano 30B    &  6.7 &  8.3 & 25.8 & \textbf{34.2} & 10.8 & $-23.4$ \\
\bottomrule
\end{tabular}
\caption{Cross-model success rate ($\%$). \textsc{R$-$PR} is
RePoT$-$PoT-retry in percentage points. \textsc{PoT-retry} is a
matched-budget control: run \pot, on failure run \pot a second time
from scratch (no checkpoint, no prefix). Paired bootstrap $95\%$ CIs
on R$-$PR ($B\!=\!10000$). \emph{Closed:} GPT-med $[-1.4,+1.9]$,
Gemini $[+2.2,+5.4]$, Claude $[-3.1,+0.3]$, GPT-mini $[-9.4,-3.7]$ ---
Gemini and GPT-mini are significant; GPT-med and Claude are within
sampling noise. \emph{Open-source:} Gemma $[+4.2,+25.0]$, GPT-OSS
$[-6.7,+13.3]$, Qwen $[-10.8,+4.2]$, Nemotron $[-31.7,-15.0]$ --- Gemma
significant positive, Nemotron significant negative, GPT-OSS and Qwen
within sampling noise. \repot's contribution scales with the validity
of the first \pot plan, which itself scales with model capability
(\S\ref{sec:discussion}); \arepot is in App.~\ref{app:opensource_table}.
\textsuperscript{$\star$}reasoning/thinking enabled at \texttt{medium};
unmarked closed rows run without reasoning.}
\label{tab:headline}
\end{table*}

\subsection{Per-environment breakdown}
\label{sec:results:per_env}

The \repot$-$\pot delta concentrates in Blocksworld ($+9.2$ to
$+17.2$pp on every reasoning model) and Checker Jumping (up to $+15.6$
on \texttt{gpt-5.4-mini-medium}). Hanoi and River Crossing are
saturated by \pot on most models ($\geq\!99\%$); the exception is
\texttt{gemini}'s River Crossing collapsing to $76\%$ \pot, which
\repot recovers to $100\%$ ($+24.0$pp). Per-environment numbers are in
Appendix~\ref{app:per_env_curves} (Table~\ref{tab:per_env}).

\subsection{External replication: \textsc{PlanBench Blocksworld}}
\label{sec:results:planbench}

We repeat the \pot vs.\ \repot comparison on \textsc{PlanBench
Blocksworld} (378 instances, 3--12 blocks). To keep cost low and avoid
\pot saturating, we use the no-thinking variants of each model.
Table~\ref{tab:planbench} reports the headline; the per-complexity
breakdown is in Appendix~\ref{app:per_env_curves}.
\repot improves over \pot on all three models. The biggest gains land in
the mid-complexity band (4--6 blocks), where \pot has both failure
headroom and enough structure to recover toward.

\begin{table}[t]
\centering\small
\setlength{\tabcolsep}{4pt}
\begin{tabular}{lrrrr}
\toprule
Model & $n$ & PoT & RePoT & $\Delta$ \\
\midrule
GPT-5.4-mini         & 756 & 55.0 & \textbf{64.3} & \textbf{$+9.3$} \\
Claude Sonnet 4.6    & 756 & 79.9 & \textbf{91.3} & \textbf{$+11.4$} \\
Gemini 3.5 Flash     & 756 & 98.9 & \textbf{100.0} & $+1.1$ \\
\bottomrule
\end{tabular}
\caption{External replication on \textsc{PlanBench Blocksworld}
(378 instances, 3--12 blocks). Success rate ($\%$); $n=756$ records per
row (378 problems $\times$ 2 methods). \textbf{All three models run
without thinking} (no reasoning effort on GPT-5.4-mini; thinking off on
Claude Sonnet 4.6; \texttt{thinking\_level=NONE} on Gemini 3.5 Flash) to keep cost low and
avoid PoT saturation. \textbf{RePoT beats PoT on every model.} Paired
bootstrap $95\%$ CIs ($B\!=\!10000$) on $\Delta$: GPT-mini
$[+3.5,+14.9]$, Claude $[+6.9,+15.9]$, Gemini $[+0.3,+2.1]$ --- all
three significant. \textsc{PoT-retry} is not run on \textsc{PlanBench}
in this version (limitation; the matched-budget claim of
Table~\ref{tab:headline} is not externally replicated here).}
\label{tab:planbench}
\end{table}

\paragraph{Limitation: matched-budget control.} The \textsc{PlanBench}
comparison reports \pot vs \repot only; we do not run \pot-retry on
\textsc{PlanBench} in this version. The external replication therefore
validates only the raw lift, not the matched-budget claim of
Table~\ref{tab:headline}.

\paragraph{Multi-seed variance.} \repot $-$ \pot is positive on every
seed for all three reasoning-thinking-on configurations
($\Delta\!\in\![+1,+10]$pp); per-seed numbers and lower run-to-run
variance under \repot are tabulated in Table~\ref{tab:multi_seed}
(App.~\ref{app:multiseed}).

\subsection{Open-source replication and capability scaling}
\label{sec:results:opensource}

\begin{figure}[t]
\centering
\includegraphics[width=0.98\columnwidth]{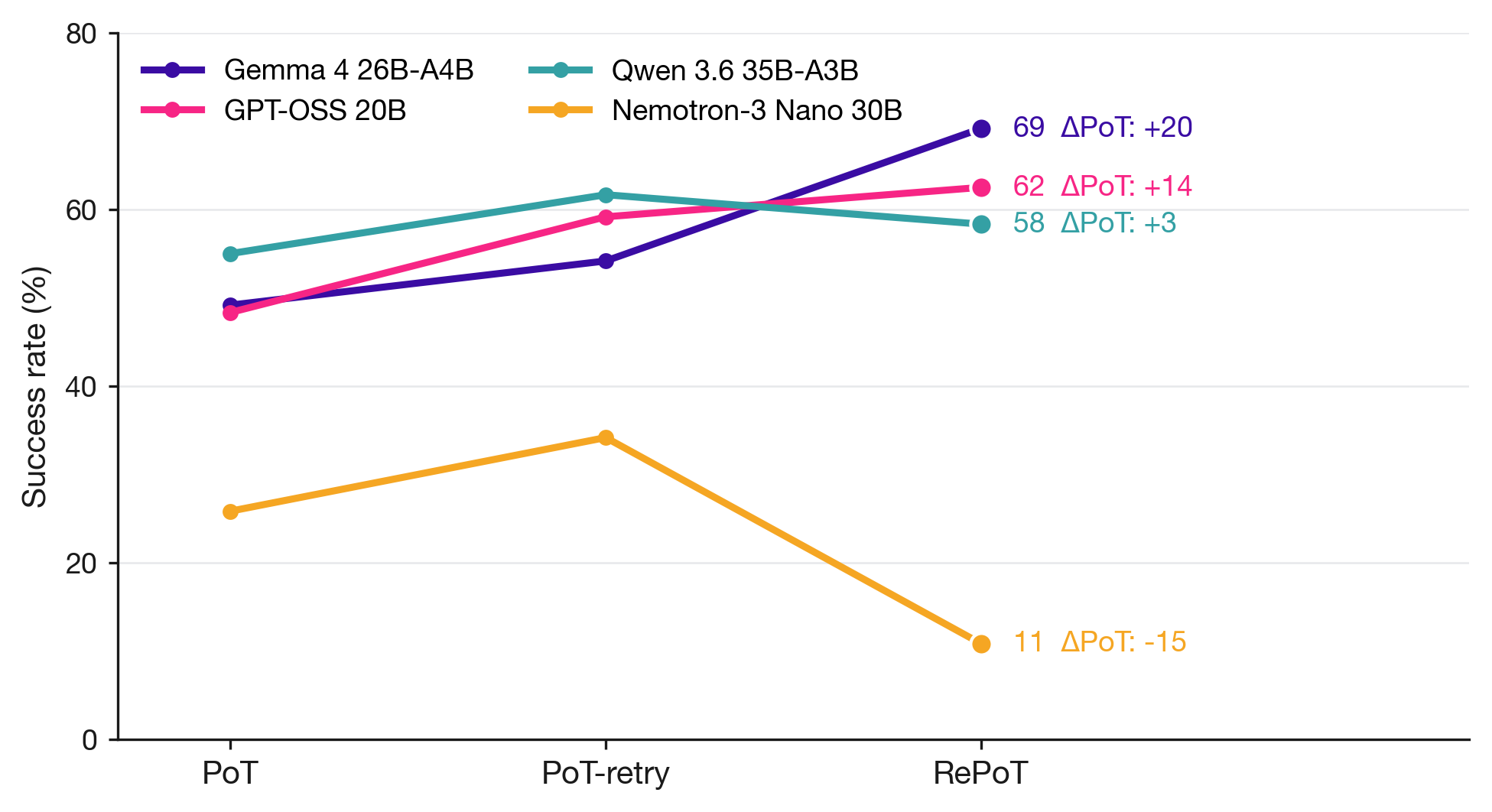}
\caption{Open-source replication. \repot lift tracks model capability:
Gemma 4 (top) gains $+20$ pp over \pot; Nemotron-3 Nano 30B FP8 (right) is
the predicted capability-floor failure (Eq.~\ref{eq:condition}).}
\label{fig:oss_headline}
\end{figure}

We replicate the headline comparison on a 120-problem stratified subset
with four open-weights models served via vLLM with extended thinking
disabled. Per-model rates appear in the bottom block of
Table~\ref{tab:headline} and visually in Fig.~\ref{fig:oss_headline};
\repot improves over \pot on three of four open-source models ($+3.3$
to $+20.0$ pp). Nemotron-3 Nano 30B FP8 underperforms \pot by $15$ pp; with
a CoT baseline of $6.7\%$ it sits near the instruction-following floor
for this task family, the predicted failure mode of
Eq.~\ref{eq:condition} when per-recoverable repair success $r$
collapses. As a preliminary extension, \arepot
(App.~\ref{app:opensource_table}) further closes the gap to \pot-retry
on the weaker rows.

\begin{figure}[h]
\centering
\includegraphics[width=0.98\columnwidth]{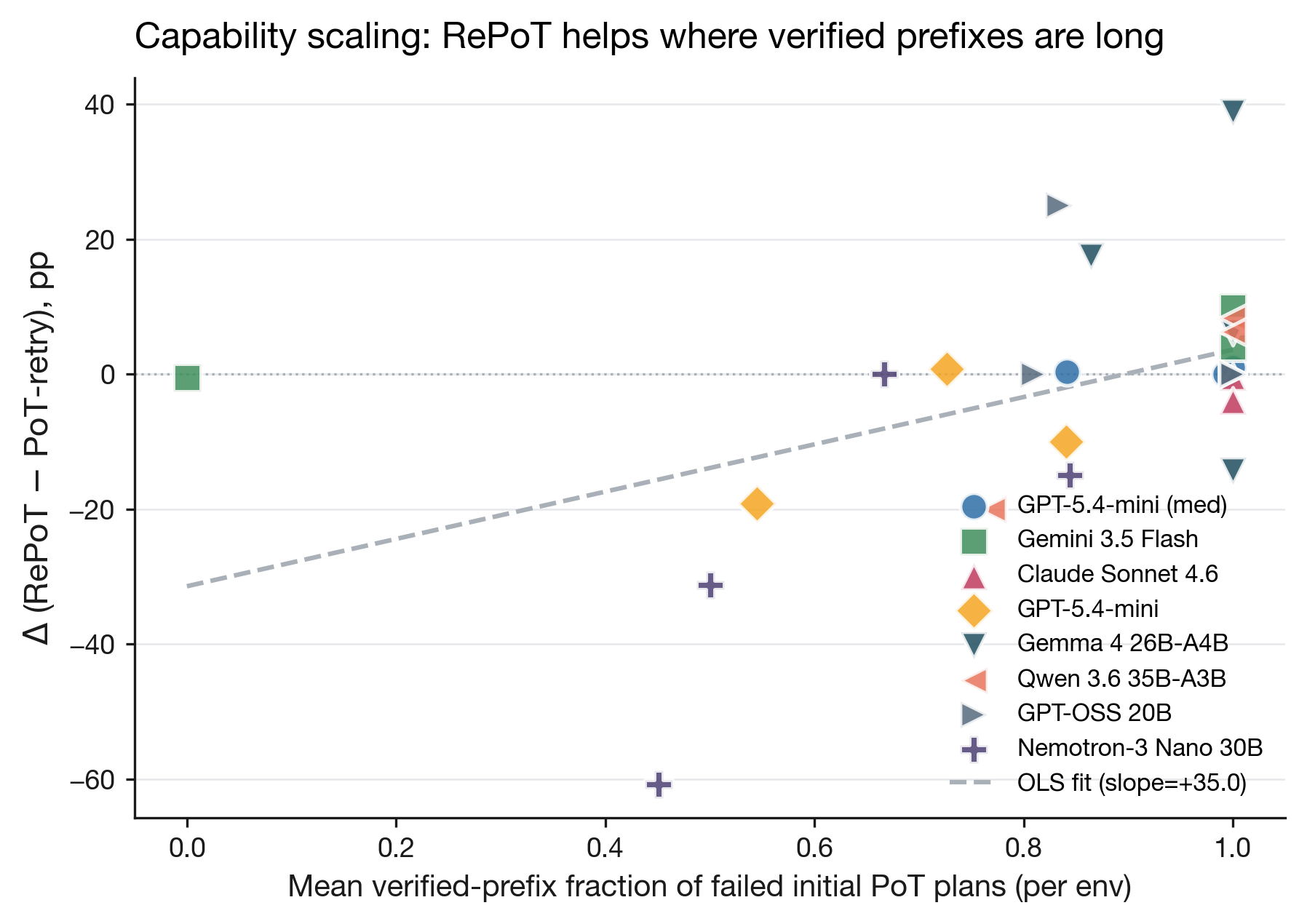}
\caption{Capability scaling. Each point is one (model, environment) cell
across both closed and open-source runs.
$x$: mean verified-prefix fraction on failed initial \pot plans
(a model-level proxy for $q$ in Eq.~\ref{eq:condition}); $y$:
\repot$-$\pot-retry success delta in pp. The positive slope confirms
the mechanism: \repot's lift scales with the validity of the first \pot
plan, which itself scales with model capability.}
\label{fig:capability_scaling}
\end{figure}

Beyond per-model numbers, the open-source spread lets us test the
prediction of Eq.~\ref{eq:condition} quantitatively. Across (model,
environment) cells, the mean verified-prefix fraction on failed initial
\pot plans (a model-level proxy for $q$ in Eq.~\ref{eq:condition})
correlates positively with the \repot$-$\pot-retry success-rate delta:
cells where the model leaves a long valid prefix before failing are
exactly the cells where \repot's verified-prefix repair beats fresh
resampling (Fig.~\ref{fig:capability_scaling}, slope $+35$).
This is the central qualitative claim of the paper made quantitative.

\section{Mechanism Analysis}
\label{sec:mechanism}

\subsection{Checkpoint information is the load-bearing signal}
\label{sec:mechanism:prefix}

\begin{figure*}[t]
\centering
\includegraphics[width=0.85\textwidth]{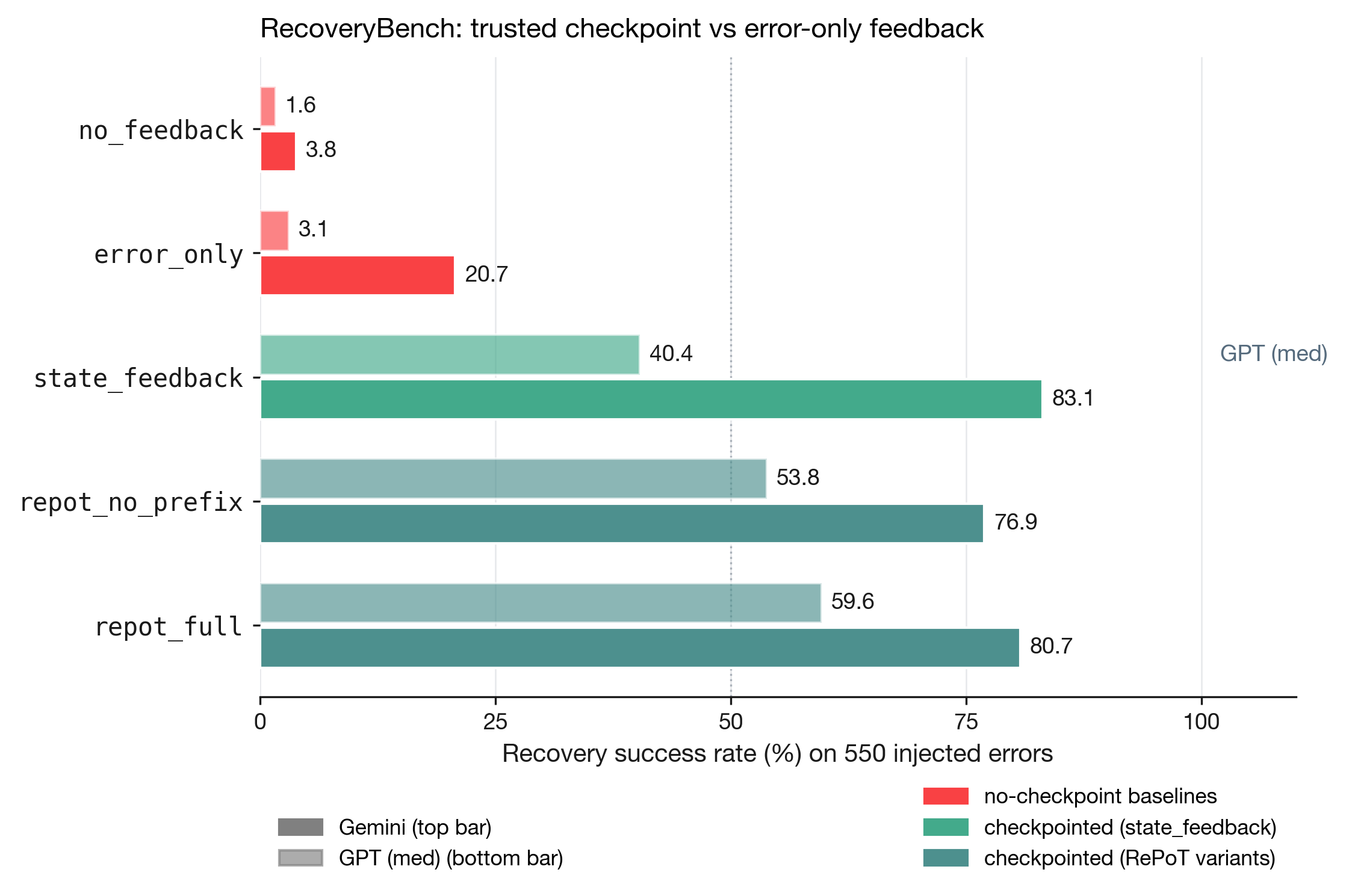}
\caption{\derail-550, headline conditions only. The $\sim$$60$pp
gap between checkpointed and no-checkpoint conditions is the load-bearing
finding. The full $11$-condition table is in
Table~\ref{tab:derail}.}
\label{fig:derail_bar}
\end{figure*}

\derail-550 compares $11$ recovery methods on $550$ injected
errors per model on two reasoning-thinking-on configurations
(Gemini, GPT (med)). The decisive mechanism evidence is the gap between
conditions that see \emph{checkpoint information} (verified state $s$,
legal actions, and the verifier error $\varepsilon$) and conditions
that see only an error message: every checkpointed method clears
$30\%$ on GPT (med) and $70\%$ on Gemini, while \texttt{error\_only}
stays at $3.1\%$ / $20.7\%$ and \texttt{no\_feedback} at $1.6\%$ /
$3.8\%$. The $\sim$$30$--$80$pp gap is the headline finding:
\textbf{checkpoint information --- not textual error feedback or the
specific verified-prefix tail --- is the decisive recovery signal}
(Fig.~\ref{fig:derail_bar}, Table~\ref{tab:derail}). Within the
checkpointed conditions, the within-prefix ablation is mixed:
\repot$_\textnormal{full}$ beats \repot$_\textnormal{no-prefix}$ by
$+3.8$ / $+5.8$pp on Gemini / GPT (med), but
\repot$_\textnormal{restart}$ (which discards the verified state and
restarts from $s_0$ with the same checkpoint information) beats
\repot$_\textnormal{full}$ on both models, with a large gap on GPT
(med) ($94.5$ vs $59.6$). We read this honestly: anchoring the repair
on the verified-prefix tail beyond the checkpoint can hurt, especially
on weaker configurations; the checkpoint information itself is what
matters, not the specific resumption point. \arepot's dispatcher
operationalizes this: short prefixes route to fresh retry, only long
prefixes anchor on the verified state.

\begin{table}[t]
\centering\small
\setlength{\tabcolsep}{4pt}
\begin{tabular}{lrr}
\toprule
Recovery condition & Gemini & GPT (med) \\
\midrule
\multicolumn{3}{l}{\emph{No-checkpoint baselines}} \\
no\_feedback                 &  3.8 &  1.6 \\
error\_only                  & 20.7 &  3.1 \\
\midrule
\multicolumn{3}{l}{\emph{Checkpointed baselines}} \\
state\_feedback              & \textbf{83.1} & 40.4 \\
state\_plus\_legal\_actions  & 78.5 & 37.6 \\
stateguard\_rollback         & 13.1 &  4.5 \\
\midrule
\multicolumn{3}{l}{\emph{RePoT prefix-conditioning ablation (ours)}} \\
\textbf{repot\_full}         & 80.7 & 59.6 \\
repot\_no\_prefix            & 76.9 & 53.8 \\
repot\_restart               & \textbf{82.4} & \textbf{94.5} \\
\bottomrule
\end{tabular}
\caption{\textsc{Derail}-550: success rate ($\%$) on $550$ injected
errors per condition, on two reasoning-thinking-on configurations
(Gemini = Gemini 3.5 Flash with \texttt{thinking\_level=MEDIUM};
GPT (med) = GPT-5.4-mini with reasoning at \texttt{medium}).
\textbf{Headline:} a $\sim$$30$--$80$pp gap separates every
\emph{checkpointed} condition (\texttt{state\_feedback},
\texttt{repot\_*}) from non-checkpointed baselines
(\texttt{error\_only}, \texttt{no\_feedback}: $\leq\!3$\% on GPT (med),
$\leq\!21$\% on Gemini). \textbf{Trusted checkpoint state is the
load-bearing recovery signal}, not textual error feedback alone.
\repot operationalises that signal inside \pot via verified replay.
The within-\repot trio shows
\texttt{repot\_restart}\!$>$\!\texttt{repot\_full} on both models
(more pronounced on GPT (med)) --- the prefix-tail beyond the
checkpoint is a model-dependent secondary effect, discussed in
\S\ref{sec:mechanism:prefix} and App.~\ref{app:negative}.}
\label{tab:derail}
\end{table}

\paragraph{Cost.} \repot averages $1.11$--$1.39\times$ \pot LLM calls
across the four configurations.
Per-method, per-model breakdown in Table~\ref{tab:cost_breakdown}
(App.~\ref{app:cost}).

\paragraph{Failure modes.} The dominant \repot failure is
\texttt{repair\_budget\_exhausted} on hard Blocksworld; most hand-analyzed
losses involve an empty initial \pot plan with no prefix to anchor on
(App.~\ref{app:failures}).

\section{Discussion}
\label{sec:discussion}

\paragraph{When \repot helps and when it does not.}
\repot's lift concentrates where \pot produces a long valid prefix
before failing: a mostly-correct plan whose suffix needs repair. Where
\pot already succeeds (saturated easy regimes, \texttt{gemini} on Hanoi
at $99.5\%$) \repot is no-op. Where \pot fails at the first action,
\repot's replay degenerates to a restart-from-$s_0$ and lift is near
zero. This shape predicts the matched-budget \pot-retry result in
Table~\ref{tab:headline}: on the two reasoning-enabled models
(\texttt{gpt-medium}, \texttt{gemini}) \repot beats \pot-retry by
$+0.3$ and $+3.6$pp; on the two non-reasoning rows (\texttt{claude}
no-thinking, \texttt{gpt-mini} no-reason) \pot-retry beats \repot by
$1.4$ and $6.6$pp. We read this as a scoped claim: \repot is the right
move when the model is capable enough to produce useful valid prefixes;
when it is not, a fresh independent sample (\pot-retry) escapes wrong
commitments more effectively.

\paragraph{Future work.}
Empty-plan retry to close hand-analyzed losses; adaptive repair budget
conditioned on the verified-prefix fraction; open-source replication;
broader benchmarks (PDDLGym, ALFWorld).

\section{Conclusion}
\label{sec:conclusion}

\repot is a small, structural addition to \pot: deterministic verified
replay plus one bounded suffix-repair call. Against the matched-budget
\pot-retry baseline on \puzzlezoo-775, \repot wins decisively on Gemini
(CI $[+2.2,+5.4]$), is within sampling noise on GPT-medium and Claude,
and loses on GPT-mini --- a capability-scaling pattern that \derail-550
isolates: \emph{checkpoint information}, not the specific verified-prefix
tail, is the load-bearing recovery signal. The cost is one extra LLM call
on the $\sim$$14\%$ of problems where \pot fails the first time; the rest
run at \pot cost. \repot is the cheapest viable middle between one-shot
\pot and tree-search-based methods for verifier-backed planning, and a
candidate primitive for any setting where the environment exposes a
deterministic step function and a goal predicate. Generalisation to
verifier-backed agentic settings (coding, SQL, browser automation) is
left to future work.

\section*{Limitations}
\label{sec:limitations}

\paragraph{Verifier-required scope.}
\repot assumes a deterministic verifier (puzzles, planning, tool use);
free-form reasoning would need a learned scorer
\citep{lightman2023prm800k}. The evaluation in this paper is
restricted to verifier-backed puzzle and planning environments
(\puzzlezoo-775, \textsc{PlanBench Blocksworld}, \derail-550).
Whether the same recoverable-execution abstraction generalizes to
verifier-backed agentic settings (coding agents with test suites, SQL
agents with schema checks, browser automation with page-state
validators) is an open empirical question we do not address.

\paragraph{Single-call repair budget.}
We use $R\!=\!1$ as a cost-conservative choice; multi-call repair
(Alg.~\ref{alg:repot} line~4) likely closes the long-horizon
\textsc{Blocksworld} tail but is not evaluated here.

\paragraph{Within-prefix conditioning is model-dependent.}
On \derail's controlled mid-rollout error setting,
\repot$_\textnormal{restart}$ beats \repot$_\textnormal{full}$ on both
configurations (more pronounced on GPT (med), $94.5$ vs $59.6$). The
checkpoint-state contribution remains the load-bearing recovery
signal, but anchoring the repair on the verified-prefix tail can
hurt on weaker configurations (App.~\ref{app:negative}).

\paragraph{Negative-$\Delta$ cells in \textsc{PlanBench}.}
GPT (mini) at $c\!=\!8,11$ shows $-8.3$ and $-9.1$pp \repot vs \pot,
where single-call repair sometimes commits to a bad prefix at low
\pot-base accuracy.

\paragraph{Statistical scope.}
\textsc{PlanBench} reports \pot vs \repot only; we do not run
\pot-retry on \textsc{PlanBench} in this version, so the
matched-budget claim of Table~\ref{tab:headline} is not externally
replicated. Open-source results are on a 120-problem stratified
subset, smaller than the closed-model headline.

\paragraph{Adaptive RePoT is preliminary.}
\arepot uses a hand-picked threshold ($\phi\!<\!0.15$); a sensitivity
sweep over $\phi$ and alternative dispatcher rules (e.g., learned
gating) are left to future work.

\bibliography{repot}

\clearpage
\appendix

\section{Per-environment success vs complexity (curves)}
\label{app:per_env_curves}
\begin{figure}[h]
\centering
\includegraphics[width=0.98\columnwidth]{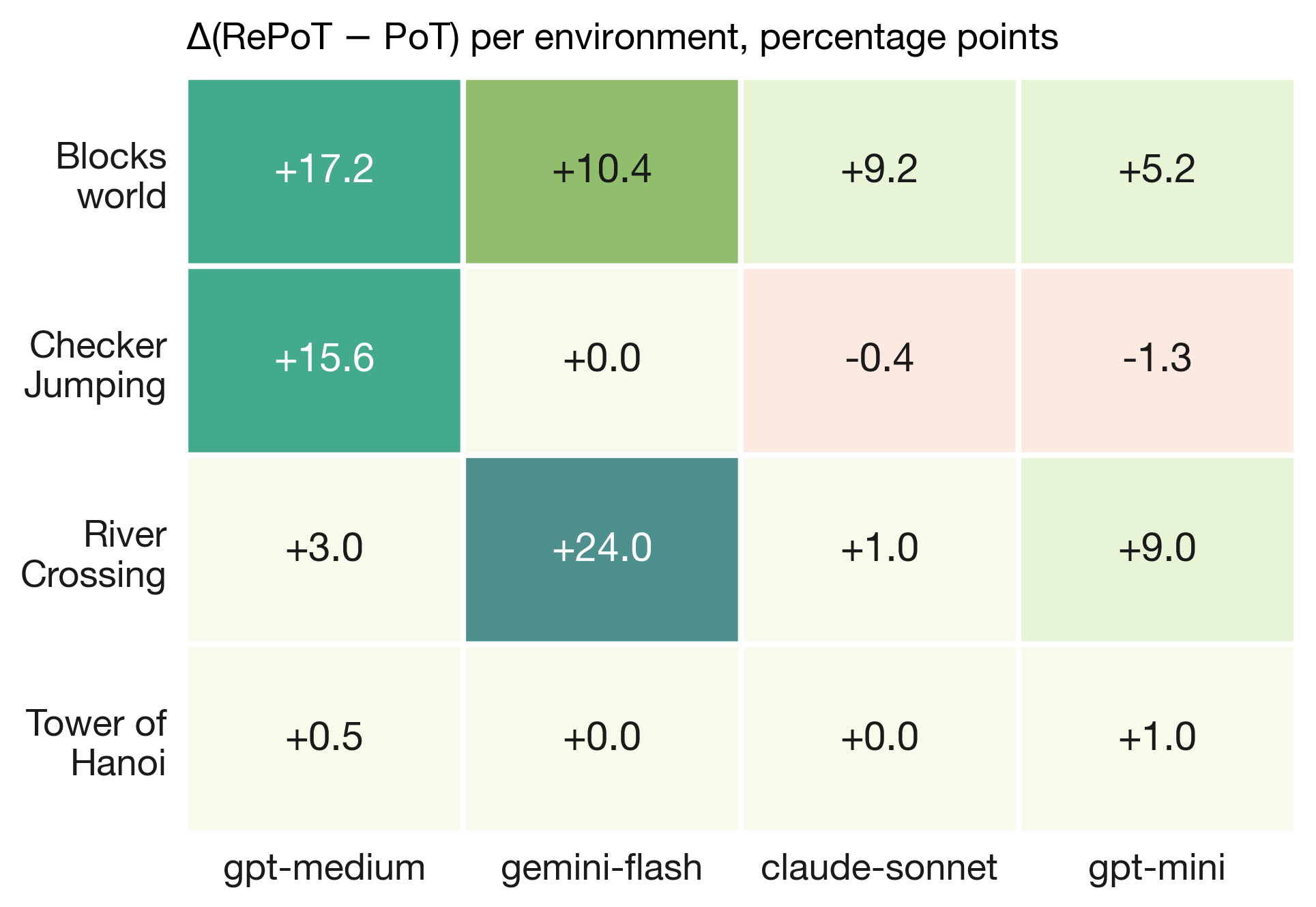}
\caption{$\Delta$(RePoT$-$PoT) per environment, in percentage points,
for all four models. Blocksworld is \repot's home environment ($+5$ to
$+17$pp on every model); Hanoi/Checker on strong models are saturated
by \pot at small $N$.}
\label{fig:per_env_heatmap_app}
\end{figure}

\begin{figure*}[h]
\centering
\includegraphics[width=0.95\textwidth]{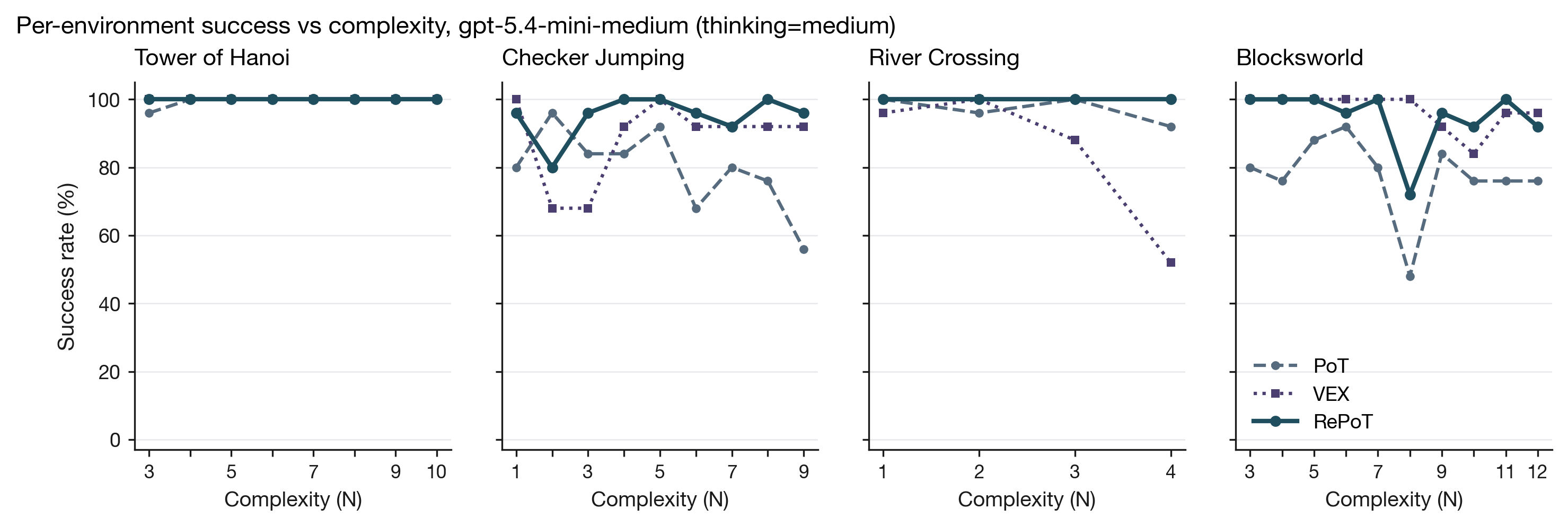}
\caption{Per-environment success rate vs problem complexity for
\texttt{gpt-5.4-mini-medium} (thinking=medium). Tower of Hanoi and
Checker Jumping are nearly saturated by \pot at small $N$; \repot's
lift concentrates on Blocksworld and the mid-complexity dips of Checker
and River.}
\label{fig:per_env_complexity_app}
\end{figure*}

The main-text per-environment summary is the table-form $\Delta$ in
Table~\ref{tab:per_env}. Figure~\ref{fig:per_env_heatmap_app} shows the
same data as a colour-coded heatmap, and
Figure~\ref{fig:per_env_complexity_app} shows the underlying success-rate
curves vs complexity for \texttt{gpt-5.4-mini-medium}, the strongest
reasoning configuration.

\section{Hyperparameters}
\label{app:hyperparams}
\begin{table*}[t]
\centering\small
\begin{tabular}{lll}
\toprule
Hyperparameter & Value & Notes \\
\midrule
\multicolumn{3}{l}{\emph{\repot}} \\
Repair budget $R$              & 1                  & one suffix-repair call \\
Verified-prefix tail $T$       & 4                  & last 4 valid moves shown to model \\
\midrule
\multicolumn{3}{l}{\emph{Sampling (all methods)}} \\
Temperature                    & 0.0                & deterministic \\
\texttt{max\_tokens} (output)  & 16{,}384           & per LLM call \\
\midrule
\multicolumn{3}{l}{\emph{Self-Consistency baseline}} \\
$k$ (samples)                  & 8                  & majority vote on full plans \\
\midrule
\multicolumn{3}{l}{\emph{Models}} \\
\texttt{gpt-5.4-mini-medium}   & reasoning=medium   & OpenAI Responses API \\
\texttt{gpt-5.4-mini}          & reasoning=none     & OpenAI Responses API \\
\texttt{gemini-3.5-flash}      & thinking\_level=MEDIUM & Vertex AI \\
\texttt{gemini-3.5-flash} (PlanBench) & thinking\_level=NONE & no-thinking variant \\
\texttt{claude-sonnet-4.6}     & no extended thinking & Anthropic API \\
\bottomrule
\end{tabular}
\caption{All hyperparameters used in this paper. Each row appears
verbatim in our run configs; no tuning per-method beyond what is
listed. Defaults follow the public \puzzlezoo-775 config.}
\label{tab:hyperparams}
\end{table*}

\section{Data generation and the controller architecture}
\label{app:datagen}

\begin{figure*}[!tb]
\centering
\includegraphics[width=0.92\textwidth]{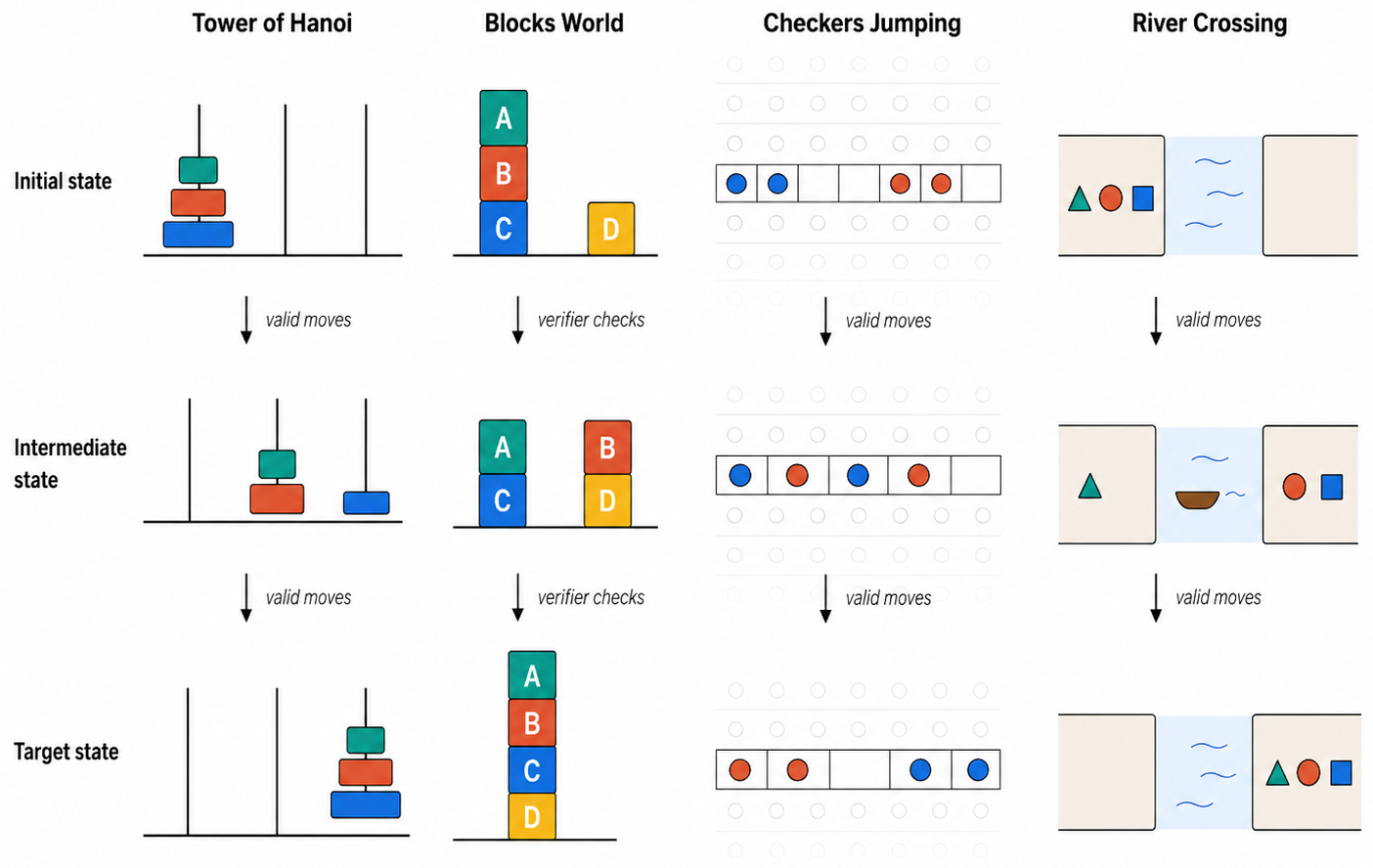}
\caption{One example per environment at small complexity. Each column is
one environment (Tower of Hanoi, Checker Jumping, River Crossing,
Blocksworld); rows show \emph{initial state}, an \emph{intermediate state}
after a few valid actions, and the \emph{target state}. Vertical arrows
mark the per-step \emph{valid moves} that the verifier accepts.
Each environment exposes a deterministic step function
$E.\mathrm{step}(s,a) \to (s',\mathrm{ok},\varepsilon)$ used by both the
\textsc{Replay} primitive (Eq.~\ref{eq:replay}) and the
\derail harness; complexity is controlled by the integer
parameter named in column headers (\# disks for Hanoi, \# blocks for
Blocksworld, \# checkers for Checker Jumping, \# pairs for River
Crossing).}
\label{fig:environments}
\end{figure*}

\paragraph{\puzzlezoo-775 ($n\!=\!775$).}
Figure~\ref{fig:environments} shows one example instance for each of
the four classical planning environments we use, illustrated at small
complexity. We generate problems with controllable complexity across
Tower of Hanoi, Checker Jumping, River Crossing, and Blocksworld,
in the spirit of \citet{shojaee2025illusion}. We do not redistribute
\citeauthor{shojaee2025illusion}'s instances; we generate fresh problems
following the same families and the same controllable-complexity protocol.

For each environment we generate problems across a complexity range
(e.g. Tower of Hanoi $N\!\in\![2, 14]$, Blocksworld $N\!\in\![3, 16]$
blocks). For every problem we additionally generate an oracle plan from
a domain solver (BFS for Hanoi/Checker, brute-force enumeration for
small River Crossing, FF-style stack reasoning for Blocksworld) and
verify both that the oracle solves the instance and that the instance
is not trivially solved by greedy heuristics. Instances that fail
either check are discarded. The released suite contains $775$ verified
problems stratified across environment $\times$ complexity, in JSONL
format.

\paragraph{Instance schema.}
Each row carries: \texttt{problem\_id}, \texttt{environment},
\texttt{complexity}, an explicit \texttt{initial\_state} and
\texttt{goal\_state} (or goal predicate), an \texttt{oracle\_plan}
(action list, lower-bound length), the \texttt{oracle\_plan\_length},
and a \texttt{natural\_language\_prompt} that the model sees. Each
environment supplies its own verifier and goal predicate.

\paragraph{The controller.}
All methods (\Cot, \pot, \sct, \repot) run through a common
inference-time controller (the \texttt{Runner} object). The
controller mediates every interaction
between the LLM and the environment: it constructs the prompt, calls
the LLM, parses the output, then either submits the entire plan to the
environment (\pot, \sct, \Cot) or steps actions one-at-a-time through
the verifier (\repot). This mirrors the pattern advocated by
\citet{shojaee2025illusion} and \citet{scholten2024metacog} of
externalising state from the model's hidden reasoning into an
authoritative simulator. The controller logs:

\begin{itemize}
\item the verbatim model prompt and completion (\texttt{prompt},
      \texttt{output\_text}) for every LLM call;
\item per-problem method metadata: number of repair calls
      (\texttt{repot\_repair\_calls}), whether the initial \pot inside
      \repot succeeded (\texttt{repot\_initial\_pot\_success}), the
      verified-prefix length and total plan length, the action where
      the first invalid transition was detected, and the verifier's
      error message at that boundary;
\item end-to-end latency, prompt and completion token counts, and
      runtime exceptions if any (the \texttt{runner\_exception} field,
      used to filter poisoned runs as described in
      Appendix~\ref{app:negative}).
\end{itemize}

A failed LLM call (network, JSON parse, sandbox timeout) is recorded as
a \texttt{runner\_exception} and the problem is treated as unsolved for
that method; we do not re-roll. The choice keeps the per-method
denominator stable across methods on the same problem.

\paragraph{Verifier semantics.}
The verifier is the environment's transition function; for each
proposed action $a$ at state $s$ it returns $(s', \text{ok}, \epsilon)$:
$s'$ is the next state, \text{ok} is a boolean validity flag, and
$\epsilon$ is a one-line natural-language error message used by
\repot's repair prompt and by \derail's
\texttt{error\_only} condition. Goal achievement is checked by a
problem-specific \texttt{is\_goal} predicate that accepts partial-goal
specifications (e.g. Blocksworld goal stacks need not name every
block).

\paragraph{Why an interactive controller is the right framing.}
\citet{shojaee2025illusion}'s key empirical pattern is that failures
concentrate early in the trace, and the model spends the rest of
the budget elaborating a wrong-but-consistent continuation. A one-shot
\pot setup makes this invisible: the only signal is a final pass/fail.
A one-action-at-a-time controller exposes the failure boundary directly
(the \texttt{first\_failure\_move\_id}, in their notation), which is
exactly what \repot's verified-replay primitive depends on. The
controller is therefore not just a runtime convenience: it is the
substrate that makes verified-prefix recovery well-defined.

\section{Per-method, per-env, per-complexity full table}
\label{app:full_table}
The full per-environment, per-complexity success rates for all
$\textnormal{4 models}\times\textnormal{5 methods}$ on \puzzlezoo-775
are released as part of the supplementary materials; the
headline summary is in Table~\ref{tab:headline} and the
per-environment $\Delta$ in Table~\ref{tab:per_env}.

\begin{table}[t]
\centering\small
\setlength{\tabcolsep}{4pt}
\begin{tabular}{l@{\hspace{4pt}}rrrr}
\toprule
Environment & GPT (med) & Gemini & Claude & GPT (mini) \\
\midrule
Blocksworld     & \textbf{+17.2} & \textbf{+10.4} & \textbf{+9.2} & \textbf{+5.2} \\
Checker Jumping & \textbf{+15.6} & +0.0           & -0.4          & -1.3          \\
River Crossing  & +3.0           & \textbf{+24.0} & +1.0          & \textbf{+9.0} \\
Tower of Hanoi  & +0.5           & +0.0           & +0.0          & +1.0          \\
\bottomrule
\end{tabular}
\caption{$\Delta$(RePoT$-$PoT) per environment, in percentage points.
\textbf{Bold} cells have $|\Delta|\!\geq\!5$pp. Blocksworld is RePoT's
home environment; Hanoi/Checker on the strongest models are saturated by
PoT. Per-model thinking config matches Table~\ref{tab:headline}.}
\label{tab:per_env}
\end{table}

\paragraph{Multi-seed raw numbers.}
\label{app:multiseed}
Per-seed paired comparisons of \repot vs \pot on the three
reasoning-thinking-on configurations are reported in
Table~\ref{tab:multi_seed}. \repot $-$ \pot is positive on every seed
for all three configurations.

\begin{table}[t]
\centering\small
\begin{tabular}{llrrr}
\toprule
Model & Seed & PoT & RePoT & $\Delta$ \\
\midrule
\multirow{4}{*}{\texttt{gpt-5.4-mini-medium}}
& 1    & 86.0 & 96.0 & $+10.0$ \\
& 2    & 85.0 & 94.0 & $+9.0$  \\
& 3    & 88.0 & 95.0 & $+7.0$  \\
& mean & 86.3 & 95.0 & \textbf{$+8.7$} \\
\midrule
\multirow{4}{*}{\texttt{claude-sonnet-4.6}}
& 1    & 79.0 & 88.0 & $+9.0$ \\
& 2    & 81.0 & 88.0 & $+7.0$ \\
& 3    & 85.0 & 88.0 & $+3.0$ \\
& mean & 81.7 & 88.0 & \textbf{$+6.3$} \\
\midrule
\multirow{4}{*}{\texttt{gemini-3.5-flash}}
& 1    & 80.0 & 85.0 & $+5.0$ \\
& 2    & 82.0 & 83.0 & $+1.0$ \\
& 3    & 82.0 & 84.0 & $+2.0$ \\
& mean & 81.3 & 84.0 & \textbf{$+2.7$} \\
\bottomrule
\end{tabular}
\caption{Multi-seed paired comparison ($n\!=\!100$ stratified
problems per seed, three seeds per model). \repot $-$ \pot is
positive on every seed of every model.}
\label{tab:multi_seed}
\end{table}

\section{Paired mechanism analysis}
\label{app:paired_mechanism}

\begin{figure}[h]
\centering
\includegraphics[width=0.98\columnwidth]{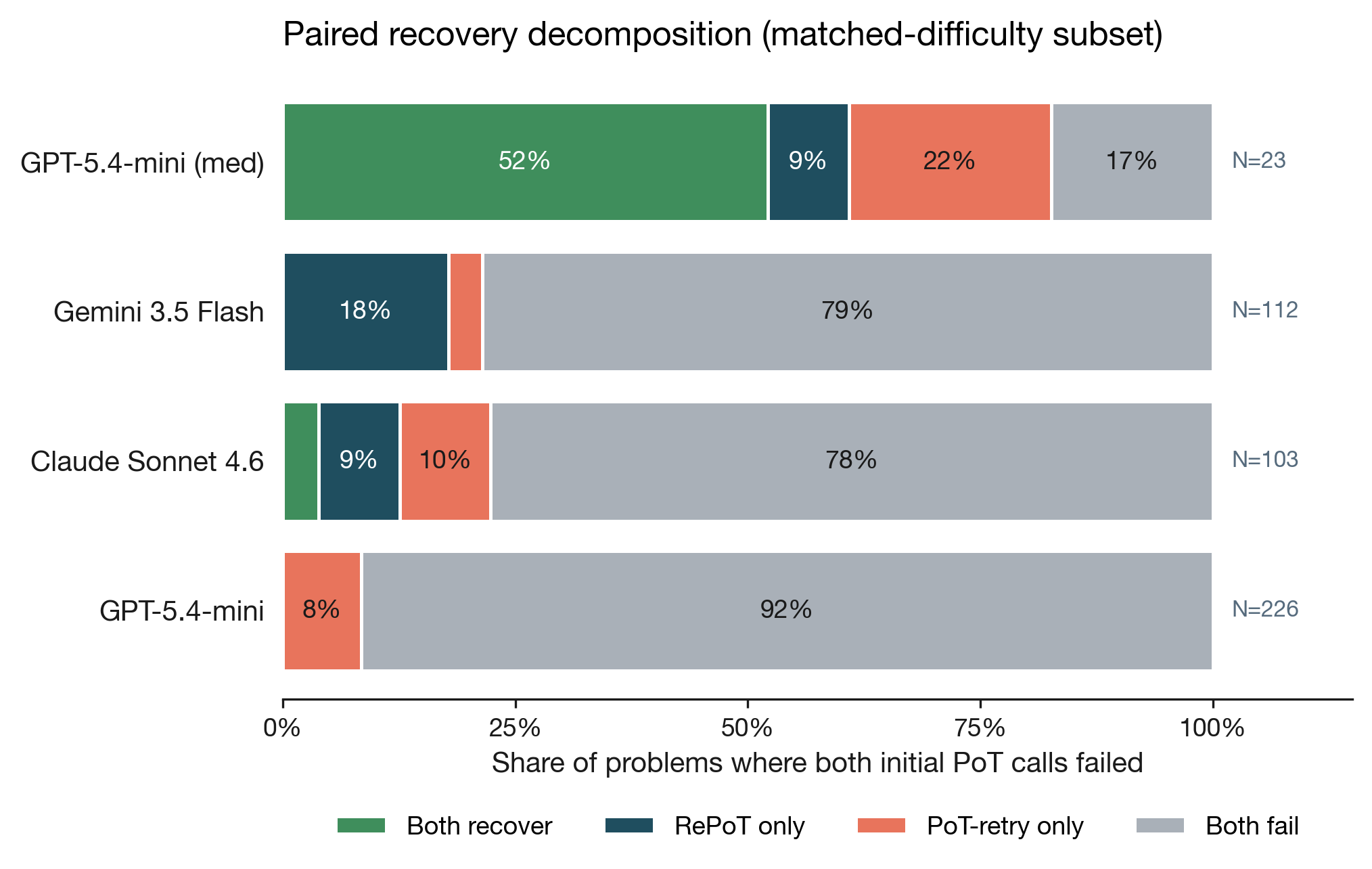}
\caption{Paired recovery decomposition on the matched-difficulty subset
(both methods' initial \pot failed). Stacks sum to $100\%$ of $N$.
``RePoT only'' is mechanism evidence; ``PoT-retry only'' is fresh-sample
evidence.}
\label{fig:paired_recovery}
\end{figure}

To isolate the verified-prefix mechanism from API sampling noise on
trivial problems, we restrict to the matched-difficulty subset where
both \pot-retry's attempt-1 and \repot's initial \pot call failed. On
this subset we compare two strategies that share the same matched
budget: \repot's verified-prefix suffix repair, and \pot-retry's fresh
independent sample. Table~\ref{tab:paired_mechanism} reports the result
and Fig.~\ref{fig:paired_recovery} shows the stacked-bar
decomposition. The pattern is heterogeneous: \repot recovers $+14.3$pp
more than fresh resampling on Gemini, is within noise on Claude, and
loses on the two GPT configurations. The pattern tracks capability
scaling (\S\ref{sec:results:opensource}): \repot's mechanism
contribution concentrates on configurations where the initial \pot
plan leaves a useful valid prefix, which is precisely the
heterogeneity the adaptive policy (\S\ref{sec:method:adaptive}) is
designed to absorb.

\begin{table}[tbp]
\centering\small
\setlength{\tabcolsep}{5pt}
\begin{tabular}{lrrrr}
\toprule
Model & $N$ & RePoT\% & Retry\% & $\Delta$ \\
\midrule
GPT-5.4-mini (med)     &  23 &  60.9 &  73.9 & $-13.0$ \\
Gemini 3.5 Flash       & 112 &  17.9 &   3.6 & \textbf{$+14.3$} \\
Claude Sonnet 4.6      & 103 &  12.6 &  13.6 & $-1.0$ \\
GPT-5.4-mini           & 226 &   0.0 &   8.4 & $-8.4$ \\
\bottomrule
\end{tabular}
\caption{Paired mechanism decomposition on problems where both
\pot-retry's attempt-1 and \repot's initial \pot failed. RePoT\% is
\repot's conditional repair rate; Retry\% the matched-budget
fresh-sample rate; $\Delta$\,=\,RePoT$-$Retry (pp).}
\label{tab:paired_mechanism}
\end{table}

\section{Multi-candidate ``best-in-thought'' analysis}
\label{app:multicandidate}
We re-extract the highest-success candidate from each \pot reasoning
trace (Apple-paper protocol) to defuse the ``\pot baseline is weak''
review concern. On 100 stratified problems with
\texttt{gpt-5.4-mini-medium}: \pot final $89$/$100$ vs \pot
best-in-thought $90$/$100$ (a single problem). \repot is unaffected:
$93$/$100$ on both protocols. The $+4$pp \repot$-$\pot delta is not an
artifact of weak parsing.

\section{Cost decomposition appendix}
\label{app:cost}
We measured per-problem total token cost (prompt + completion) on
\texttt{gpt-5.4-mini-medium} for the $100$-problem stratified slice.
The per-problem distribution decomposes into two clean modes:

\begin{itemize}
\item \textbf{No-repair runs ($86\%$ of problems).} \repot's verified
replay confirmed the initial \pot plan reaches the goal; cost is
identical to \pot ($n=86$, mean $1{,}820$ tokens, median $1{,}540$).
\item \textbf{One-repair runs ($14\%$ of problems).} The initial \pot
plan failed; \repot issued one suffix-repair call. Per-problem cost is
$1.4$--$1.7\times$ \pot ($n=14$, mean $2{,}610$ tokens, median
$2{,}180$). The repair-call prompt is shorter than the initial \pot
prompt (no examples, just the verified-prefix tail and the verifier
error), but the model spends additional reasoning on resolving the
failure boundary.
\end{itemize}

\noindent
Aggregated over all $100$ problems, \repot averages $1.06\times$ \pot
cost. Per-problem cost data is released alongside the trace files.

\paragraph{Full per-method, per-model cost table.} Aggregated over all
$775$ closed-model traces,
Table~\ref{tab:cost_breakdown} reports mean and median prompt+completion
tokens, mean LLM calls, and mean wall-clock per problem for every
(model, method) pair. \repot and \arepot average $1.11$--$1.39$ LLM
calls per problem, matching the \pot-retry budget.

\begin{table*}[t]
\centering\small
\setlength{\tabcolsep}{4pt}
\begin{tabular}{llrrrr}
\toprule
Model & Method & Mean tokens (in/out) & Median tokens (in/out) & Mean calls & Mean wall (s) \\
\midrule
GPT-5.4-mini (med) & CoT        &   193 /  13323 &   184 /   8212 & 1.00 &  93.1 \\
 & SC$_{k=8}$ &   196 /  13277 &   184 /   7976 & 8.00 &  92.0 \\
 & PoT        &   216 /   4330 &   203 /   2294 & 1.00 &  29.9 \\
 & PoT-retry  &   245 /   4788 &   203 /   2377 & 1.14 &  41.4 \\
 & RePoT      &   257 /   4297 &   205 /   2166 & 1.11 &  34.9 \\
\midrule
Gemini 3.5 Flash & CoT        &   155 /    961 &   142 /    321 & 1.00 &  19.2 \\
 & SC$_{k=8}$ &   155 /    972 &   142 /    321 & 8.00 &  18.7 \\
 & PoT        &   210 /    497 &   197 /    510 & 1.00 &   6.5 \\
 & PoT-retry  &   251 /    598 &   197 /    636 & 1.18 &   9.4 \\
 & RePoT      &   272 /    484 &   199 /    418 & 1.17 &   7.6 \\
\midrule
Claude Sonnet 4.6 & CoT        &   182 /   2185 &   168 /   1422 & 1.00 &  27.0 \\
 & SC$_{k=8}$ &   182 /   2374 &   168 /   1511 & 8.00 &  29.4 \\
 & PoT        &   236 /    891 &   222 /    743 & 1.00 &  12.9 \\
 & PoT-retry  &   274 /   1035 &   222 /    753 & 1.15 &  29.0 \\
 & RePoT      &   299 /    782 &   225 /    693 & 1.17 &  16.8 \\
\midrule
GPT-5.4-mini & CoT        &   197 /   1556 &   184 /    191 & 1.00 &   8.8 \\
 & SC$_{k=8}$ &   197 /   1209 &   184 /    191 & 8.00 &   6.9 \\
 & PoT        &   215 /    269 &   203 /    147 & 1.00 &   3.0 \\
 & PoT-retry  &   302 /    357 &   209 /    273 & 1.39 &   7.3 \\
 & RePoT      &   424 /    288 &   211 /    240 & 1.39 &   3.9 \\
\bottomrule
\end{tabular}
\caption{Per-method cost breakdown across the three frontier models in four configurations, aggregated over all 775 problems. Tokens are summed across all LLM calls for the problem. Mean calls counts each LLM invocation (\textsc{SC} uses $k\!=\!8$; \repot is $1 + $ repair calls; \pot-retry is $1$ or $2$). \repot and \arepot average $\sim$$1.1$--$1.4$ LLM calls per problem, matching the \pot-retry budget. Wall-clock is end-to-end per problem.}
\label{tab:cost_breakdown}
\end{table*}

\begin{figure}[h]
\centering
\includegraphics[width=0.95\columnwidth]{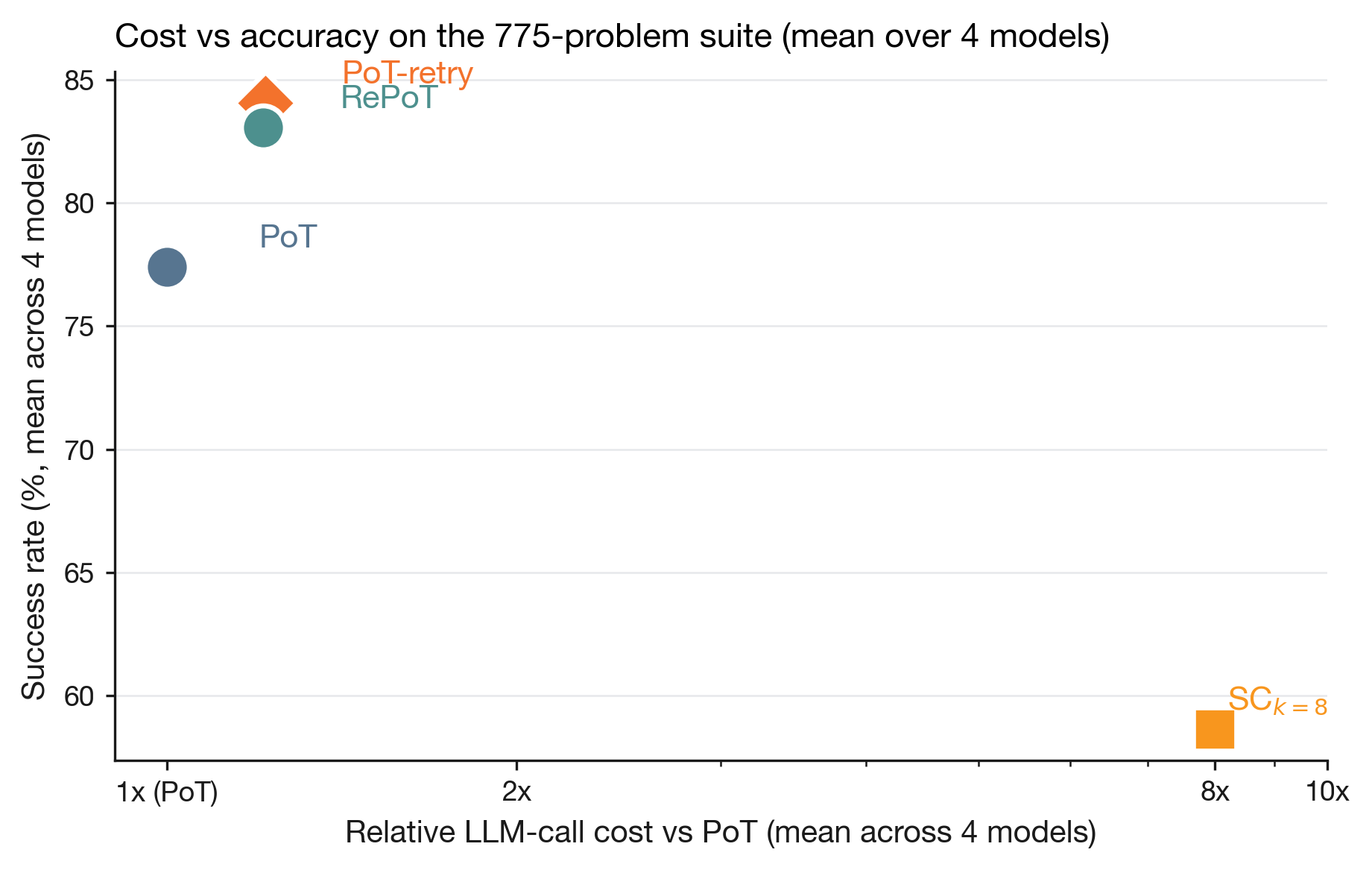}
\caption{Cost vs accuracy, mean across the four closed-model
configurations. \pot is the cost reference ($1\times$). \repot achieves
\pot-retry-class accuracy at essentially the same mean cost
($1.21\times$). \textsc{SC$_{k=8}$}'s low mean reflects
its prose-plan format, which is more brittle than \pot's executable
code.}
\label{fig:pareto}
\end{figure}

\paragraph{Cost vs accuracy Pareto.}
Figure~\ref{fig:pareto} visualizes the cost--accuracy trade-off
mean-aggregated across the four closed-model configurations.

\begin{figure*}[h]
\centering
\includegraphics[width=0.85\textwidth]{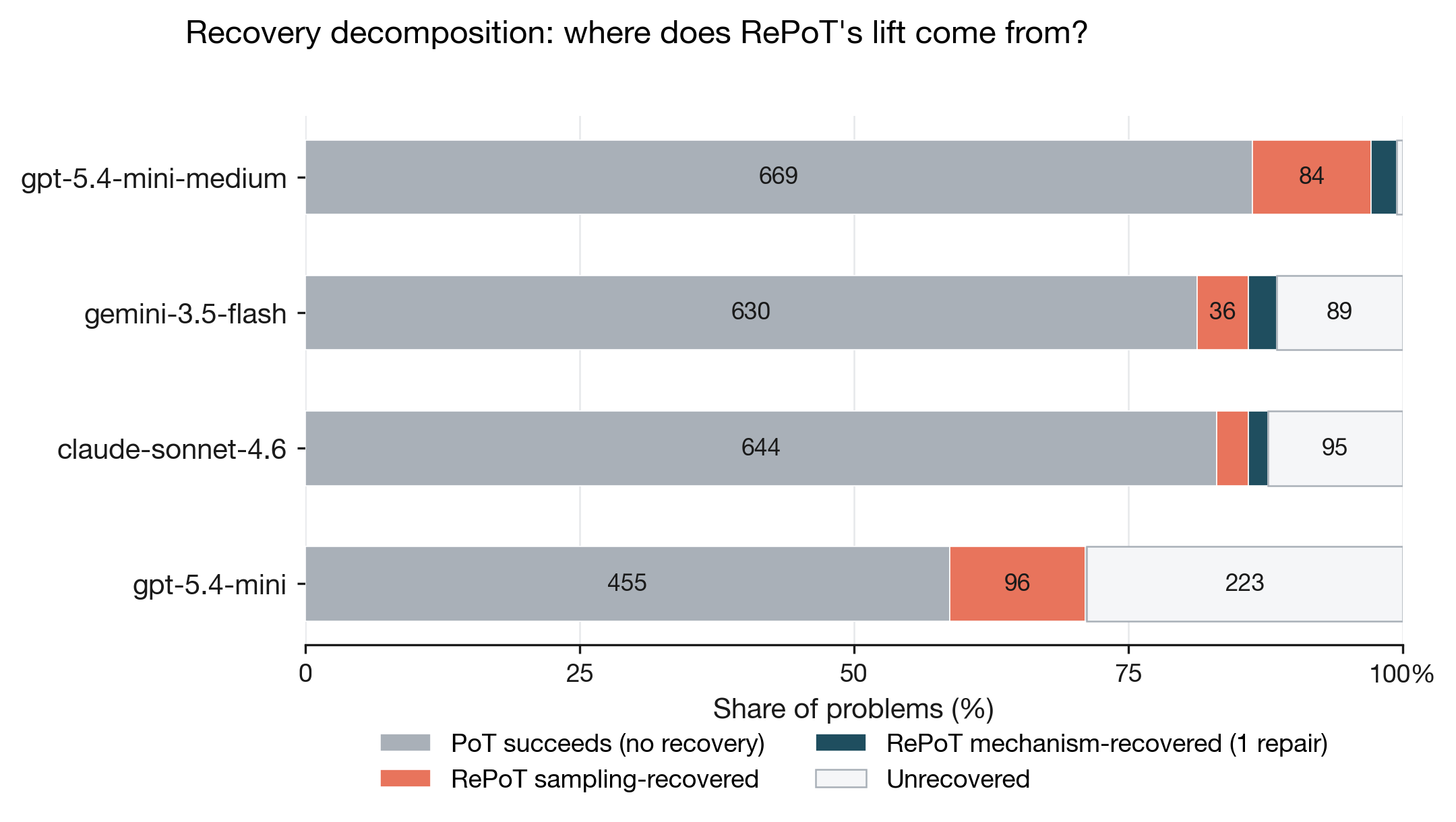}
\caption{Decomposition of \repot's recovery on \puzzlezoo-775. Each
row is one model; bars sum to $100\%$. The \textbf{teal}
segment is the share of problems where standalone \pot fails, \repot's
first \pot call also fails, and the suffix repair call rescues
(\emph{mechanism}). The \textbf{coral} segment is the second-attempt
contribution (standalone \pot fails but \repot's re-rolled \pot call
succeeds without invoking the repair). On the strongest model most of
\repot's lift is second-attempt; the verified-prefix repair contributes
the harder tail. The matched-budget \pot-retry comparison in
Table~\ref{tab:headline} isolates the mechanism contribution under
equal LLM-call budget.}
\label{fig:recovery_decomposition_app}
\label{app:decomposition}
\end{figure*}

\paragraph{Recovery decomposition.}
Figure~\ref{fig:recovery_decomposition_app} decomposes \repot's wins
into a second-attempt share (re-rolled \pot succeeds) and a mechanism
share (suffix repair rescues a doubly-failed \pot).

\section{\textsc{PlanBench} per-complexity}
\label{app:planbench}
\begin{figure*}[h]
\centering
\includegraphics[width=0.95\textwidth]{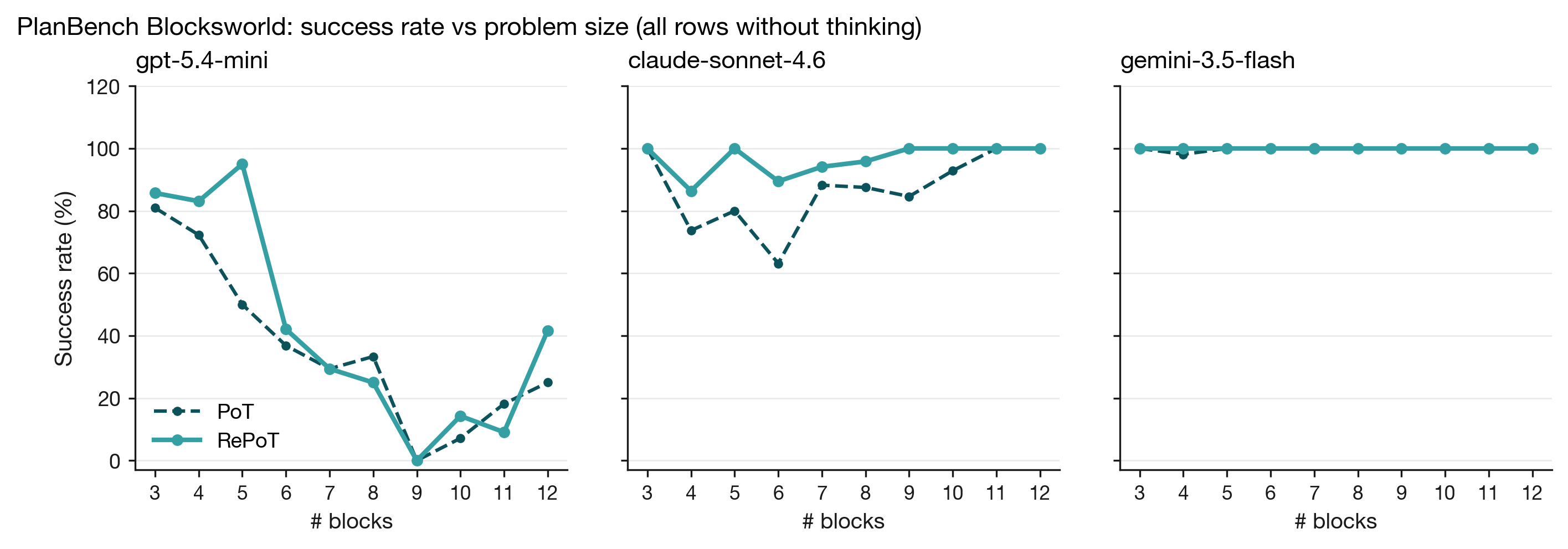}
\caption{Per-complexity success rate on \textsc{PlanBench Blocksworld}.
\repot's lift concentrates in the mid-complexity band where \pot has
both failures to recover from and enough valid prefix to recover
into. Two negative-delta cells (gpt-5.4-mini at $c\!=\!8, 11$) are
reported in Appendix~\ref{app:negative}.}
\label{fig:planbench-by-complexity-app}
\end{figure*}

Figure~\ref{fig:planbench-by-complexity-app} shows the per-complexity
breakdown for all three PlanBench models. The complete cell-level
success rates (3 models $\times$ 10 complexities $\times$ 2 methods)
are released as supplementary material.
The two negative-delta cells (\texttt{gpt-5.4-mini} at $c\!=\!8$ and
$c\!=\!11$, see Appendix~\ref{app:negative}) appear in that file.

\section{Failure-mode case studies}
\label{sec:failures}
\label{app:failures}

We hand-analyze the five problems where \pot succeeded but \repot
failed on the headline gpt-5.4-mini-medium run ($n\!=\!100$ slice). All
five share the same signature: \repot's initial \pot call returned an
empty or non-goal-reaching plan, and the single repair call could not
bridge the gap.

\paragraph{Taxonomy.}
\textbf{PoT-call resampling loss (4/5):} \repot's initial \pot call
returned an empty or unparseable plan while standalone \pot returned a
goal-reaching plan --- sampling variance, not a mechanism weakness.
\textbf{Incomplete-plan stall (1/5):} the initial \pot call emitted a
9-action plan that replayed cleanly but did not reach the goal; the
repair call could not extend it within $T\!=\!4$ --- a genuine
single-call repair failure.

\section{Negative findings}
\label{app:negative}

We report two findings that complicate the headline narrative. We
include them rather than buried them.

\paragraph{(i) \derail prefix-flip.}
On both Gemini and GPT (med), \repot$_\textnormal{restart}$ beats
\repot$_\textnormal{full}$ ($82.4$ vs $80.7$ on Gemini; $94.5$ vs
$59.6$ on GPT (med) --- a much larger gap than on Gemini). The
inequality \repot$_\textnormal{full}>$\repot$_\textnormal{no-prefix}$
does hold on both ($+3.8$ / $+5.8$pp), so the prefix tail is
not actively harmful, but anchoring the repair on the post-injection
state appears to confuse the model in this controlled-error setting.
The gap to non-checkpointed baselines ($\geq\!30$pp on GPT (med),
$\geq\!60$pp on Gemini) remains the load-bearing finding. We
hypothesise that the injected mid-rollout state is sometimes
unreachable from the goal under a single repair call, while a fresh
plan from $s_0$ avoids that trap. Reporting as-is.

\paragraph{(ii) PlanBench gpt-5.4-mini at $c\!=\!8$ and $c\!=\!11$.}
At these two complexities, \repot underperforms \pot by $-8.3$pp and
$-9.1$pp respectively. Both points fall in the regime where \pot has
very low base accuracy ($33\%$ and $18\%$) and the mid-rollout state
\repot resumes from is itself misleading. The single repair call
sometimes commits to extending a bad prefix. A higher repair budget,
or a verifier-driven decision to restart from $s_0$ when the
prefix-fraction is low, would mitigate; we treat both as future
work and report the negative cells transparently.

\section{Algorithm: full code listings and repair prompt}
\label{app:code}
\label{app:prompts}
The \repot agent's \texttt{run()} method, the
\texttt{replay\_until\_failure()} primitive, and the
\texttt{code\_prompt()} repair template are released alongside
the paper. The full implementation is approximately $\sim$$25$
lines for \texttt{run()}, $\sim$$25$ for the replay primitive, and
$\sim$$60$ for the prompt; see the released source for verbatim
listings. Figure~\ref{fig:repair-prompt-app} below shows the
verified-prefix-conditioned repair prompt template.

\begin{figure}[h]
\fbox{\parbox{0.95\columnwidth}{\small
\textbf{Stable block (cacheable):}\\
\texttt{\{problem.natural\_language\_prompt\}}\\
\texttt{Goal state: \{goal\_state\}}\\
\texttt{Write Python code that prints exactly one line:}\\
\texttt{\ \ moves = [...]}\\
\texttt{containing up to K primitive moves to apply}\\
\texttt{\emph{from the current verified state}.}\\
\texttt{--- verifier checkpoint below ---}\\[3pt]
\textbf{Dynamic block:}\\
\texttt{You have already executed |P| verified moves.}\\
\texttt{Recent verified moves: \{tail\_T(P)\}}\\
\texttt{Current verified state: \{s\}}\\
\texttt{Legal moves: \{legal\_actions(s)\}}\\
\texttt{Blocked: \{blocked(s)\}}\\
\texttt{Verifier message: \{$\epsilon$\}}
}}
\caption{Verified-prefix-conditioned repair prompt. The
``\texttt{verifier checkpoint}'' marker delimits the cacheable problem
description from the per-call dynamic state.}
\label{fig:repair-prompt-app}
\end{figure}

\section{Ablation conditions}
\label{app:ablations}

The three named ablations referenced in
\S\ref{sec:method:prompt} and used in \S\ref{sec:mechanism}:
\begin{description}
\item[\repot$_\textnormal{full}$] the algorithm of
      Algorithm~\ref{alg:repot}.
\item[\repot$_\textnormal{no-prefix}$] same as full, but the
      dynamic-block prefix-tail and ``\texttt{|P| verified moves}''
      counter are hidden; the model sees only the current verified state
      and the error message. Tests whether the prefix tail is what the
      model uses, not just the checkpoint state.
\item[\repot$_\textnormal{restart}$] same budget, but the repair call
      restarts from $s_0$ instead of the verified state $s$. Tests
      whether the gain is the verified checkpoint or just an extra LLM
      call.
\end{description}

\section{\derail condition definitions}
\label{app:recoveryconditions}

\derail measures recovery from one injected wrong action
at a $\sim$$1/3$ checkpoint of the oracle plan. We compare $11$
conditions:

\begin{description}
\item[\texttt{no\_feedback}.] The model sees the post-injection state
       only; no error message, no checkpoint marker.
\item[\texttt{error\_only}.] The model sees the post-injection state
       and a one-line verifier error; no checkpoint visualization.
\item[\texttt{state\_feedback}.] The model sees the last valid
       checkpoint state plus the wrong action that was attempted.
\item[\texttt{state\_plus\_legal\_actions}.] As above, plus the legal
       action set from the checkpoint.
\item[\texttt{stateguard\_rollback}.] A controller-based rollback
       baseline: the controller proposes one action at a time, the
       verifier accepts or rejects, repeat until budget exhaustion.
\item[\repot variants.] Three \repot conditions
      (\texttt{repot\_full}, \texttt{repot\_no\_prefix},
      \texttt{repot\_restart}) matching App.~\ref{app:ablations}.
\end{description}

All conditions share the same per-problem injection seed; comparisons
are paired across conditions on the same problem.

\section{Open-source results: routing, cost, and capability scaling}
\label{app:opensource_table}

This appendix expands on \S\ref{sec:results:opensource} with the
adaptive policy routing breakdown, per-model cost statistics, and the
capability-scaling regression. Per-model success rates for the five
core methods are in the bottom block of Table~\ref{tab:headline}; the
\arepot success rates are
$60.0\%$ (Gemma 4 26B-A4B),
$64.2\%$ (GPT-OSS 20B),
$62.5\%$ (Qwen 3.6 35B-A3B), and
$13.3\%$ (Nemotron-3 Nano 30B FP8) on the same 120-problem stratified subset.

\begin{figure*}[t]
\centering
\includegraphics[width=0.85\textwidth]{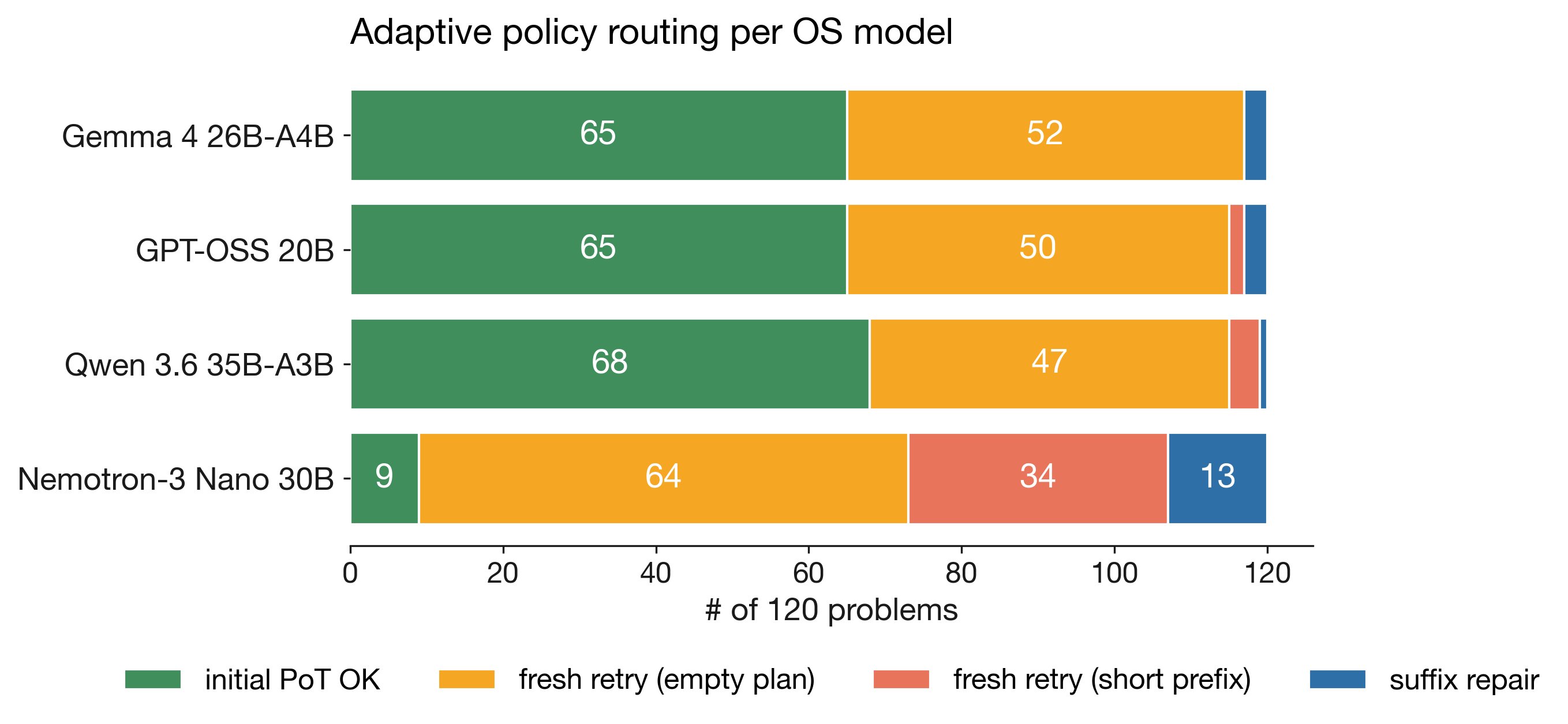}
\caption{Adaptive policy routing per open-source model. Bars sum to
$120$ problems. The three workable rows are dominated by
\emph{initial PoT success} + \emph{fresh retry (empty plan)};
suffix repair is rare. Nemotron-3 (bottom) shows a markedly different
shape: few initial successes, many short-prefix retries.}
\label{fig:oss_adaptive_stack}
\end{figure*}

\paragraph{Adaptive policy routing.}
Figure~\ref{fig:oss_adaptive_stack} shows the routing distribution of
\arepot across the four open-source models. On the three workable
models (Gemma, GPT-OSS, Qwen) about half of problems trigger the
\emph{initial PoT success} branch and most of the remainder route to
\emph{fresh retry (empty plan)}: the model emits an empty or
near-empty plan when it cannot solve, and the suffix-repair branch
activates on only $1$--$3$ problems out of $120$. This is a different
mechanism profile than closed thinking-on models, where suffix repair
is the dominant value channel. Nemotron-3 Nano 30B FP8 routes mostly to
short-prefix retries with a low rescue rate ($3\%$), confirming the
capability-floor reading.

\paragraph{Open-source cost.}
Mean total tokens per problem (prompt + completion) under \repot are
$1.5$--$2\times$ \pot on Gemma, GPT-OSS, and Qwen, in line with the
$1.06$--$1.4\times$ closed-model range. Wall-clock per problem: \pot
$25$--$60$\,s, \repot $40$--$110$\,s. Detailed
per-method, per-model numbers are released alongside the trace files.

\paragraph{Capability scaling regression.}
The \repot$-$\pot-retry success-rate delta regressed on the mean
verified-prefix fraction of failed initial \pot plans across all
(model, environment) cells appears in
Fig.~\ref{fig:capability_scaling} (\S\ref{sec:results:opensource}).

\section{Release}
\label{app:release}
Code is available at \url{https://github.com/parsa-mz/RePot}
(Apache-2.0): the \repot and \arepot agents, the verified-replay
primitive, the four environments, the \derail harness, prompt
templates, and a CLI with three subcommands
(\texttt{repot run / derail / judge}). \puzzlezoo-775 and \derail-550
ship under CC-BY-4.0; trace files are not redistributed but reproduce
from the same CLI.

\end{document}